\documentclass[aps,superscriptaddress,groupedaddress,nofootinbib]{revtex4}
\usepackage{natbib}
\usepackage[utf8]{inputenc}
\usepackage{physics}
\usepackage{amsmath}
\usepackage{amssymb}
\usepackage{siunitx}
\usepackage{graphicx}
\usepackage{xcolor}
\usepackage[capitalize]{cleveref}
\usepackage[english]{babel}
\usepackage[autostyle]{csquotes}

\MakeOuterQuote{"}

\newcommand{\Ket}[1]{\ket{#1}}
\newcommand{\Bra}[1]{\bra{#1}}
\newcommand{\pTr}[2]{\mathrm{Tr}_{#1} \left[ #2 \right]}
\newcommand{\avg}[1]{\textrm{E}\left\{#1\right\}}

\newcommand{\red}[1]{{#1}}

\usepackage[normalem]{ulem}
\newcommand\bluestrikeout{\bgroup\markoverwith
{\textcolor{blue}{\rule[.5ex]{2pt}{0.4pt}}}\ULon}
\begin{document}

	
	
	\title{Finite-key analysis for memory-assisted decoy-state quantum key distribution}
	\author{Guillermo \surname{Currás Lorenzo}}
	\email{G.J.CurrasLorenzo@leeds.ac.uk}
	\affiliation{School of Electronic and Electrical Engineering, University of Leeds, Leeds, UK} 
	\author{Mohsen Razavi}
	\affiliation{School of Electronic and Electrical Engineering, University of Leeds, Leeds, UK}
	\date{\today}

	\begin{abstract}
		Memory-assisted quantum key distribution (MA-QKD) systems are among novel promising solutions that can improve the key-rate scaling with channel loss. By using a middle node with quantum storage and measurement functionalities, they offer the same key-rate scaling with distance as a single-node quantum repeater. However, the distance at which they can surpass the nominal key rate of repeaterless systems, in terms of bits per second, is typically long, owing to the efficiency and/or interaction time issues when one deals with quantum memories. This crossover distance can be a few hundred kilometres, for instance, when one relies on the exchange of infinitely many key bits for the key-rate analysis. In a realistic setup, however, we should account for the finite-key effects in our analysis. Here, we show that accounting for such effects would actually favour MA-QKD setups, by reducing the crossover distance to the regime where realistic implementations can take place. We demonstrate this by rigorously analysing a decoy-state version of MA-QKD, in the finite-key regime, using memory parameters already achievable experimentally. This provides us with a better understanding of the advantages and challenges of working with memory-based systems. 
	\end{abstract}

	\maketitle

	\section{Introduction}
	
	Quantum key distribution (QKD) has made a lot of progress as part of the solution package for secure communications in the quantum era \cite{pirandola2019advances}. But, when it comes to long distances, quantum technologies still have a long way to go before they can replicate the same functionalities that public-key cryptography offers. In terrestrial networks, such as the infrastructure that today's Internet is based on, the biggest challenge to overcome is perhaps the exponential growth of loss in optical fibres \cite{gisin2015far}. This makes it extremely difficult to perform QKD at long distances without trusted middle nodes. Quantum repeaters are potential solutions, but none of their theoretical architectures can currently be implemented experimentally to the full effect \cite{muralidharan2016optimal}. For instance, probabilistic quantum repeaters \cite{duan2001long, sangouard2011quantum, piparolongdistance2015} would require quantum memory (QM) modules with high coupling efficiencies to light and with coherence times exceeding the transmission delays, which are hard to achieve together \cite{panayi2014memory}. That said, even if the current QMs are not sufficiently advanced for quantum repeaters, they may still be used to offer key-rate improvements in some of the existing QKD systems. Working on such memory-assisted QKD (MA-QKD) systems paves the way for future scalable quantum repeaters. This work studies the secret key rate for decoy-state MA-QKD systems in the practical regime where only a finite block of data is exchanged among QKD users.
	
	MA-QKD setups \cite{panayi2014memory,abruzzo2014measurement} are based on the measurement-device-independent QKD (MDI-QKD) protocol \cite{lo2012measurement}, in which Alice and Bob send BB84-encoded pulses to a middle node, Charlie, who performs a Bell-state measurement (BSM). In MDI-QKD, a raw key bit can be generated if both pulses survive the channel loss in the same round and the BSM is successful. In MA-QKD, however, Charlie employs two QMs to store the quantum state of the users' pulses, and only performs the BSM when both memories have been loaded. This will allow the pulses that arrive in different rounds to be combined to produce a key bit. {Thus, the key-rate scaling is improved from $\eta^2$ in MDI-QKD to ${\eta}$ in MA-QKD \cite{panayi2014memory}, where $\eta$ is the transmittance of the channel between Alice/Bob and Charlie. Together with the recently introduced twin-field QKD (TF-QKD) \cite{lucamarini2018overcoming}, MA-QKD is a strong contender to beat the current rate versus distance records in QKD. Such an advantage has recently been demonstrated experimentally using  silicon vacancy centres \cite{bhaskar2020experimental}.}
	
	Offering advantage in a realistic setup that relies on imperfect QMs is not without its own challenges. For instance, photon-memory coupling can introduce additional loss in the setup. Some memories have also a long photon-memory interaction time that requires users to employ a low source repetition rate. The better scaling with channel loss can only offset these effects after a certain distance, which we refer to as the crossover distance. If this distance happens to be long, it would then be difficult to experimentally implement a stable system that benefits from such an advantage. Other effects, such as decoherence in the QMs, also need to be taken into account when evaluating system performance \cite{panayi2014memory} and they typically exacerbate the situation. Additionally, in realistic setups, we should consider the effect of using weak laser pulses by the users in conjunction with finite-key effects. In this work, we develop a security analysis that accounts for all the above, and, in particular, quantify the interplay between the crossover distance and other parameters of the system. 
	
	Several analyses of MA-QKD have already been carried out, under varying assumptions and for different implementations of QMs. However, most of them \cite{abruzzo2014measurement, piparo2017multiple, piparo2017nitrogen} assume single-photon sources, which are difficult to attain in practice. In many QKD experiments, attenuated laser sources are used, instead. The multi-photon components in the signals generated by these sources introduce security loopholes, and they need to be dealt with \cite{gllp}. The decoy-state method \cite{ma2005practical} is often used to bound the leaked information from these multi-photon signals, thus closing the loophole. This method involves the statistical estimation of channel probabilities, based on data collected from the use of different laser intensities. This statistical characterisation of the channel would only be perfect if one could collect an infinite amount of data by using the channel infinitely many times. In practice, a QKD experiment will run for a fixed amount of time, and a finite-size dataset will be generated \cite{zhang2017improved}. By using statistical analyses based on concentration inequalities, it has been shown that a bound on the leaked information can be computed \cite{curty2014finite,zhang2017improved}, thus a secret key can still be distilled, with a failure probability that can be made arbitrarily small. However, as the total number of signals exchanged (the block size) gets smaller, the obtainable secret key rate is reduced. In fact, if the block size is too small, no secret key rate may be obtained at all.
	
	In this paper, we provide the first analysis of a decoy-state MA-QKD setup that  accounts for the statistical fluctuations that arise from generating a finite-size key. Previous work \cite{panayi2014memory} on MA-QKD has only considered the asymptotic limit in which the users exchange an infinite number of signals, and under simplified assumptions on the loading of QMs with attenuated laser sources.  In our finite-key analysis, we compare MA-QKD performance with that of a no-memory MDI-QKD system, by using parameters from state-of-the-art experiments on quantum memories \cite{piparo2017multiple}. We find that MA-QKD is inherently more resistant to finite-key effects, and it experiences a lower reduction in secret key rate than MDI-QKD. In particular, we see that once these effects are considered, the distance from which MA-QKD offers an advantage is reduced. This would make it easier for experimentalists to implement a decoy-state MA-QKD setup that outperforms, in terms of secret key rate versus distance, the equivalent decoy-state BB84 or MDI-QKD setups. 
	
	\red{In terms of key rate, MA-QKD may not outperform the recently introduced TF-QKD, at least with state-of-the-art quantum memories. However, one should be careful when comparing systems that have different requirements. For instance, the single-photon interference of TF-QKD demands phase stability over long channels, which is experimentally difficult, and which MA-QKD does not need. We believe that comparing MA-QKD with MDI-QKD is the fairest when it comes to the requirements of each system. We note that there exists some recent work on memory assisted TF-QKD \cite{schmidt2019memory}, which specifies under what circumstances adding quantum memories to TF-QKD setups can be advantageous. Moreover, we believe that MA-QKD is of special interest as is the very first step toward building memory-based quantum repeaters. Unlike TF-QKD, or other no-memory systems, these offer a scalable solution for long distance quantum communications. Any practical progress with quantum repeaters would be based on fully understanding and implementing MA-QKD as the simplest memory-based repeater system. Our findings for MA-QKD systems suggest that memory-based quantum repeaters may also be resilient to finite-key effects, at least when users access them with decoy-state sources.}
		
	The rest of the paper is organised as follows. In \cref{sec-systemdescription}, we describe the analysed setup, placing an emphasis on the QM modules, and the different parameters that are used for modelling them. In \cref{sec:keyanalysis}, we explain how different system parameters affect the secret-key rate. In \cref{sec:numericalresults}, we compare the secret key rate achievable in decoy-state MA-QKD and decoy-state MDI-QKD with examples from warm vapour and cold atomic ensembles. Section \ref{sec-conclusions} concludes the paper with our interpretation of the results. 

	\section{System description}
	\label{sec-systemdescription}
	
	In this section, we describe our MA-QKD setup and the assumptions we make on different devices and components of the system.
	
\begin{figure}[htbp]
				\includegraphics[width=0.7\textwidth]{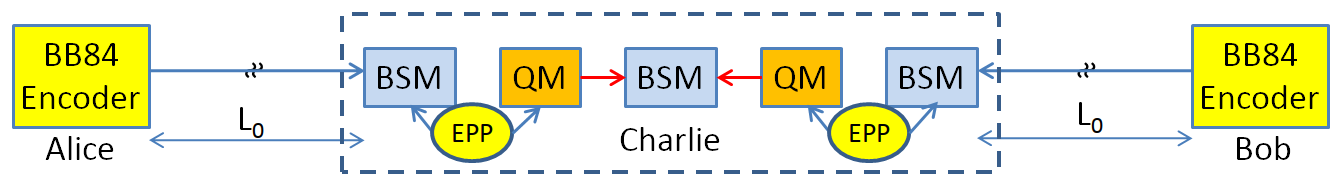}
				\caption{The schematic of an MA-QKD system. The two users Alice and Bob use decoy-state BB84 encoders to generate polarisation/phase encoded signals. Charlie, in the middle, uses entangled photon pair (EPP) sources to teleport the state sent by the users to the corresponding memories. When both memories are loaded, their states are converted back to photons and combined in the middle BSM. For an example of the BSM module, see \cref{fig:WCPSchematics} of Appendix A. }	\label{fig:schematics}
	\end{figure}

	Figure \ref{fig:schematics} shows the schematic of the MA-QKD setup considered in this work. Here, in each round, Alice and Bob each send decoy-state BB84 states in their chosen basis. Charlie verifies the receipt of the transmitted signal by generating an entangled photon pair (EPP) on each side to effectively teleport the state of the users to a local photon on his site. The side BSMs in Fig.~\ref{fig:schematics} would herald the success of such an event, in which case the remaining photon of the EPP source will be written to the corresponding QM. That is, its photonic state is transferred to the memory, and will be kept there until the state of the other user is also successfully received and teleported to its respective QM. At this point, the two QMs will be read, i.e., their states will be transferred to photons on which the middle BSM is performed. At the end of the protocol, Charlie announces his measurement results, and Alice and Bob would follow with conventional steps for sifting and post-processing of their key bits.

	Note that the teleportation scheme used here to herald and transfer the state of photons is not an ideal one. In an ideal teleportation setting, the users have to send ideal single photons, whereas here they are using weak laser pulses. The effect of the multi-photon components has then to be taken into account. We analyse the memory-loading procedure for weak laser pulses in Appendix \ref{app-memoryloading}. In this scheme, we are also delaying the writing of the second photon of the EPP until we learn about the success of teleportation. While there is a chance that the transfer of this photonic state to the QM may fail, this delayed writing process has the advantage that the QM initialisation is not necessary in each round \cite{piparo2017multiple}, but only when a writing procedure has been attempted. This helps with maximising the repetition rate of the protocol especially when the initialisation phase is time consuming. We account for the failure in transferring a local single photon to the memory by the memory writing efficiency parameter. 
	
	Finally, while, in practice, an ideal EPP source as assumed here may not be realistic, it would help us obtain the key features of our finite-key analysis without overly complicating the calculations. The former issue can be managed by techniques introduced in Ref.~\cite{piparo2017multiple}, where they propose a quasi-EPP scheme based on single-photon sources, instead. It is also possible to create a photon-QM entangled pair in certain QMs \cite{piparo2017nitrogen, Keller_trapped_ion}. In all cases, we should be careful with the possible multiple excitations we may locally create at Charlie's node to not violate the conditions for the proper operation of MA-QKD systems \cite{piparo2015ensemble,piparo2017multiple}. Under above considerations, we believe that the main result from our paper, i.e., the resilience of the decoy-state MA-QKD to finite-key effects, should still hold.

In the following, we describe the key components of our system in more detail.

\subsection{Quantum memories}

	We model QMs using a few relevant parameters to our setup, while keeping our model as general as possible:
	
	\begin{itemize}
		\item The writing efficiency, denoted as $\eta_w$, is the probability of successfully transferring a single-photon state to the quantum memory. We refer to this process by the term ``loading''.
		\item The reading efficiency, denoted as $\eta_r$, is the probability to transfer the qubit state stored in the QM back to a single photon. We assume that, at time $t$ after loading, $\eta_r (t) = \eta_{r0} \exp [-t/T_1]$, where $\eta_{r0}$ denotes the reading efficiency at time $t = 0$ and $T_1$ is the decay time constant of the QM. 
		\item The QM decoherence time constant is denoted by $T_2$. We consider two decoherence processes: dephasing and depolarisation. In the case of dephasing, for an initial state $\rho(0)$ of the QM, the state at a time $t$ after loading will be 
		\begin{equation}
		\label{eq:dephasing}
		\rho(t) = p(t) \rho(0) + [1-p(t)] \sigma_z \rho(0) \sigma_z,
		\end{equation}
		where $p(t) = [1+\exp(-t/T_2)]/2$. Dephasing will only affect $X$-basis states. {For a depolarisation process, we assume 
		\begin{equation}
		\label{eq:depol}
		\rho(t) = p(t) \rho(0) + \frac{1-p(t)}{3}[ \sigma_z \rho(0) \sigma_z +  \sigma_x \rho(0) \sigma_x + \sigma_y\rho(0) \sigma_y].
		\end{equation}
		In both cases, we treat the QM state as a qubit for which $\sigma_x$, $\sigma_y$, and $\sigma_z$ are its corresponding Pauli operators.} 
		\item We denote the interaction time with single photons as $\tau_{\textrm{int}}$, for both reading and writing procedures. We denote the initialisation time of the QM as $\tau_{\rm init}$. Because of our delayed-writing assumption, a writing procedure will always be followed by a reading procedure, and the QM only needs to be initialised after reading.
		\item The writing time is denoted as $\tau_w$, and the reading time is denoted as $\tau_r$. For our delayed writing procedure, we assume $\tau_w = \tau_{\textrm{int}}$ and $\tau_r = \tau_{\textrm{int}} + \tau_{\rm init}$. { We effectively neglect the required time for measurement in both cases.}
		\item We denote as $\tau_p$ the pulse duration of both the user sources and the EPP sources, which are assumed to have matching pulse shapes. We assume $\tau_p = \tau_w$ to maximize the writing efficiency into the memory. The MA-QKD system is to be run at a repetition rate of $R_s = 1/\tau_p$.
			\end{itemize}

	\subsection{Channel and source model}
	Similarly, we present our assumptions on the channel and the users sources:
	
	\begin{itemize}
		\item We assume that the user sources produce phase-randomised coherent states, and that the intensity of the pulse can be perfectly tuned in each round. The users select a random intensity, in terms of mean number of photons, from the set $\{z, w_1, w_2, v\}$ with probability $\{p_z, p_{w_1}, p_{w_2}, p_v\}$. Emissions with the $z$ intensity will be encoded in the $Z$ basis, and they will be used to generate the raw key. Emissions with any other intensity will be encoded in the $X$ basis, and they will be used to estimate the single-photon counts and their corresponding phase-error rate. We will refer to $z$ as the signal intensity, and to $\{w_1, w_2, v\}$ as the decoy intensities.
		Our model can work with either polarisation or phase encoding. 
		We denote the source repetition rate as $R_s$.
		\item We assume non-resolving detectors with efficiency $\eta_d$ and 
		a dark count rate $\gamma_{dc}$. The latter includes intrinsic effects as well as background photons in the channel. The dark count probability per detector per round of the protocol is $p_{\mathrm{dc}} = \gamma_{dc} \tau_p$.
		\item We denote the total length of the channel separating Alice and Bob by $L$. We assume that the central node is located exactly halfway between the users. We denote the attenuation length of the channel by $L_{\textrm{att}}$. The transmission coefficient for each leg of the channel is given by $\eta_{\mathrm{ch}} = \exp({\frac{-L}{2 L_{\textrm{att}}}})$.
		\item We consider the effect of setup misalignment between the user sources and the measurement devices in the central node. The standard way to model misalignment in QKD is by a misalignment probability $e_{\rm mis}$, and previous analyses of MA-QKD have also modelled it that way \cite{panayi2014memory}. However, as explained in Appendix \ref{app-memoryloading}, such a model is not directly applicable when considering the indirect loading of QMs with weak laser pulses. {Here, we model misalignment by assuming that the encoding modes, e.g., horizontal and vertical polarisations, have been rotated from their ideal settings by a random angle $\theta$. We then average over $\theta$ to find parameters of interest.}
		\item In our setup, we allow for the usage of frequency converters to match the frequency of the telecom signals sent by the users with that of the EPP source. The EPP source, in one leg, should generate a beam that interacts with the QM. For a degenerate EPP source, this would typically require us to downconvert the frequency of the other beam to the telecom band. One can, in principle, design a non-degenerate EPP source, but we should then be careful with the extent of multiple excitations in the source \cite{piparo2015ensemble}. We account for the efficiency of frequency converters by including additional loss in our setup.
		\end{itemize}

	\section{Key-rate analysis}
	\label{sec:keyanalysis}
	
	In this section, we find the secret key generation rate for our decoy-state MA-QKD setup, in both the asymptotic and finite-key regimes. We assume the nominal mode of operation in which no eavesdropper is present, and the system is only affected by device imperfections. Also, for simplicity, we assume that the sources used by Alice and Bob, and the channels connecting them to the middle node are identical.
	
	\subsection{Asymptotic case}
	
	In this subsection, we calculate the key rate obtainable in the limit that the users exchange an infinite number of signals. In this regime, we can assume that the signal intensity is used with probability $p_z \simeq 1$, and that the decoy-state analysis provides a perfect estimate of the single-photon channel probabilities. Under these assumptions, the secret key rate is lower bounded by \cite{panayi2014memory}
	\begin{equation}
	\label{eq:keyrateasymp}
	R \red{\geq} R_s \left[Q_{11}^{Z} \left(1-h(e_{\rm ph})\right) - fQ_{Z} h(e_{Z})  \right],
	\end{equation}
	where $Q_Z$ is the probability of generating a sifted key bit per round of the protocol, and $e_Z$ is the error rate of the sifted key. Also, $Q_{11}^{Z}$ is the single-photon contribution to $Q_{Z}$, and $e_{\rm ph}$ is the phase-error rate of these single-photon components. 
	
	Our objective here is to calculate what Alice and Bob would observe in a nominal experiment for directly measurable parameters $Q_Z$ and $e_Z$, and their corresponding estimation for $Q_{11}^{Z}$ and $e_{\rm ph}$ after using the decoy state method. For this, we mainly use the method introduced in \cite{panayi2014memory}, but we adjust it as needed to account for the specific components of our model. In particular, in the case of weak laser pulses at the source, we need to pay special attention to the modelling of misalignment in the channel. We also extend the results of \cite{panayi2014memory} to depolarising channels.
	
	Appendix \ref{app-memoryloading} provides a detailed and self-contained description of our analysis. In short, we first obtain the exact expression for loading probability $p_{\rm load}^\mu$ and loading error rate $e_{\rm load}^\mu$ when Alice/Bob sends a phase-randomised coherent state with intensity $\mu$ under a generic model for channel misalignment. This parameter would then allow us to calculate the average number of rounds needed to load both memories, and the corresponding state of the memories after a heralded loading. We will then account for memory decoherence and decay processes and calculate the rate of success, and the corresponding error rate, for the middle BSM. Section \ref{sec:AppAsymp}, in Appendix \ref{app-memoryloading}, provides the analytical form for all parameters needed in \cref{eq:keyrateasymp}.

	\subsection{Finite-key regime}
	
	Now, we calculate the secret key rate in the more realistic scenario where the number of signals exchanged by the users is finite. In this regime, we still derive the secret key from the data points for which both users have used the $Z$ basis, but we also need to take into account the rounds in which the users employ decoy intensities. In this case, we can no longer assume that the decoy-state analysis provides a perfect estimate of the single-photon statistics $Q_{11}^{Z}$ and $e_{\rm ph}$. Instead, we use a statistical analysis to bound them. Under our new assumptions, the total secret key length $K$ satisfies
	\begin{equation}
	\label{eq:finitekeyrate}
	K \geq M^{Z}_{11}[1-H(e_{\rm ph})] - M_{Z} H(e_{Z}),
	\end{equation}
	where $M_{Z}$ is the length of the sifted key, generated from the events in which both users selected the $Z$ basis (i.e., the $z$ intensity), and $e_{Z}$ is its bit error rate; $M^{Z}_{11}$ is the number of bits in this sifted key that originated from single-photon emissions, and $e_{\rm ph}$ is their phase-error rate. 
	
	 In an experimental implementation of the protocol, the measurable observables available to us are the sets $\{M^{ab}\}$ and $\{E^{ab}\}$, where $M^{ab}$ is the total number of measurement counts when Alice has used intensity $a$ {\em and} Bob has used intensity $b$, while $E^{ab}$ is the number of such events that result in error. The objective of Alice and Bob is to use this data to obtain statistical bounds on $M^{Z}_{11}$ and $e_{\rm ph}$. 
	 
	The full description of our statistical analysis appears in Appendix \ref{app-finitekey}.  We use the idea in \cite{zhou2016making} to perform our statistical fluctuation analysis using $X$-basis data only. This would make our statistical estimation procedure more efficient. By applying tight multiplicative Chernoff bounds \cite{zhang2017improved}, we are then able to use the measured counts $M^{ab}$ and $E^{ab}$ to set linear constraints on the possible values that $M^{Z}_{11}$ and $e_{\rm ph}$ could take. These constraints enable us to express the desired bounds on these quantities as the solution to two linear programs. We use the analytical estimation procedure introduced in \cite{curty2014finite} to solve these programs.
	 
	For our numerical simulations, we still need to make some assumptions on the obtained measurement results in a nominal experiment. For this purpose, we use the expected values for relevant parameters using the corresponding probability in the asymptotic regime, derived in the previous subsection. That is, we assume
	\begin{equation}
	\label{eq:eabmain}
		M^{ab} = N  Q^{ab} \mbox{\ \ and \ \ }
	E^{ab} = e_{ab} M^{ab},
		\end{equation}
	where $N$ is the total number of rounds, i.e., the number of transmitted pulses by Alice/Bob, in the protocol, $Q^{ab}$ is the probability of having a successful measurement originating from intensities $a$, for Alice, and $b$, for Bob, and $e_{ab}$ is the probability that this measurement results in an error. Section \ref{Sec:AppFinite}, in Appendix \ref{app-memoryloading}, provides the derivation and the analytical form for all these parameters. 
	
	%
	%
	%
	%
	
	
    
    \red{In our finite-key analysis, we have only considered the effect of statistical fluctuations on parameter estimation.} Thus, in our key rate formula in \cref{eq:finitekeyrate}, we have neglected some of the less significant terms that usually appear in a rigorous finite-key analysis. The latter is to adhere to the universal composable framework \cite{ben2005universal,renner2005universally}; e.g., we direct the reader to Eq.~(1) of \cite{curty2014finite}. We have neglected these terms for simplicity, as they are, in practice, only on the order of tens of bits, and because their effect is identical for the memory-assisted and no-memory systems, which the present work aims to compare.

	\section{Numerical results}
	\label{sec:numericalresults}
	In this section, we use the results of \cref{sec:keyanalysis} to simulate the secret key rate that can be achieved with the decoy-state MA-QKD scheme in \cref{fig:schematics}, in both the asymptotic and finite-key regimes. We use two types of memories for our analysis: Warm vapour atomic ensembles, which often offer high bandwidth, hence high repetition rates, but a rather low coherence time; and cold atomic ensembles, which are often slower but benefit from longer coherence times.  \cref{table:memories} summarises the relevant memory parameters used in our simulation based on the experimentally reported values in \cite{camacho2009four}, for warm vapours, and  \cite{yang2016efficient}, for cold atomic ensembles. In our simulations, we have assumed $T_1 = T_2$. 
		
	\begin{table}[h]
		\begin{tabular}{llll}
			\hline 
			& WV \cite{camacho2009four} & CA \cite{yang2016efficient}  & \red{SV} \cite{bhaskar2020experimental}   \\
			\hline 
			Writing-reading efficiency, $\eta_{w} \eta_{r0}$ & $0.05$  & $0.76$  & $0.423$  \\
			\hline 
			Decay time, $T_1$    & $\SI{120}{\micro\second}$  & $\SI{220}{\milli\second}$ & 200 $\mu$s \\
			\hline 
			Interaction time, $\tau_{\textrm{int}}$   & $\SI{1.43}{\nano\second}$ & $\SI{240}{\nano\second}$  & 142 ns \\
			\hline 
			Repetition rate, $R_s$ & $\SI{518}{\mega\hertz}$   & $\SI{4.2}{\mega\hertz}$ & 7.04 MHz\\
			\hline 
		\end{tabular}
		\caption {Parameter values of recently demonstrated warm vapour (WV) and cold atom (CA) ensembles \cite{piparo2017multiple}, as well as silicon vacancy (SV) centres, used in the simulations in this work. For simplicity, in our simulations, we assume $T_2 = T_1$.}
		\label{table:memories}
	\end{table}
	
	We compare the MA-QKD system with a no-memory MDI-QKD setup, run at a repetition rate of $\SI{1}{\giga\hertz}$, as a reference point, and study how finite-key effects change the crossover distance under different circumstances. Section \ref{Sec:AppMDI} in Appendix \ref{app-memoryloading} provides the analytical expressions used for simulating the MDI-QKD system. MDI-QKD is the closest no-QM system to MA-QKD, which enables us to make this comparison as fair as possible. They both offer measurement-device-independent features and they can both be run with minimal requirements on the source or channel phase stabilisation. The latter property is needed for advanced twin-field QKD systems, whose rate-versus-distance scaling is similar to MA-QKD, but are expected to offer higher rates if properly implemented \cite{Maeda2019, lorenzo2019tight, TF-QKD509km}.
	
	In all cases, we use the system parameters listed in \cref{table:channelparam}, which are attainable by today's technologies \cite{marsili2013detecting}. In all graphs, we optimise over the values of the intensities $\{z, w_1, w_2\}$, and assume a vacuum intensity of $v = 0.5 \cdot 10^{-3}$, since the optimal value $v=0$ may be difficult to achieve in practice. We also optimise over their selection probabilities $\{p_z, p_{w_1}, p_{w_2}, p_v\}$. {In our finite-key analysis, we assume a failure probability of $\varepsilon = 0.5\cdot 10^{-11}$ for each of the concentration bounds used in \cref{app-finitekey}; the total failure probability of the estimation process is $20 \varepsilon = 10^{-10}$.}
	
	\begin {table}[h]
	\begin{tabular}{lll}
		\hline 
		& Attenuation length of the channel, $L_{\textrm{att}}$ &  $\SI{22}{\kilo\meter}$ \\
		\hline 
		& Detector efficiency, $\eta_d$  &  $93\% $\\
		\hline
		&Detector dark count rate, $\gamma_{dc}$&  {$\SI{1}{count\per\second}$}\\
		\hline
		&{Misalignment error probability, $e_{\rm mis}$}&  {$0.5 \%$}\\
		\hline
		&\red{Conversion efficiency, $\eta_{c}$} &  \red{$0.5,1$}\\
		\hline
	\end{tabular}
	\caption {System parameter values used for the simulations in this work. {For no-memory MDI-QKD, we assume that the channel misalignment, in their respective leg of the channel, flips the state sent by each user with probability $e_{\rm mis}$. For MA-QKD, we assume that channel misalignment rotates the states sent by the users by an angle $\theta$ that follows a uniform distribution of width $2\sqrt{3e_{\rm mis}}$; see \cref{eq:misalign-equiv}, and the explanation preceding it.} } \label{table:channelparam}
	\end{table}
	
\begin{figure}[htbp]
		\centering
		\includegraphics[width=0.48\textwidth]{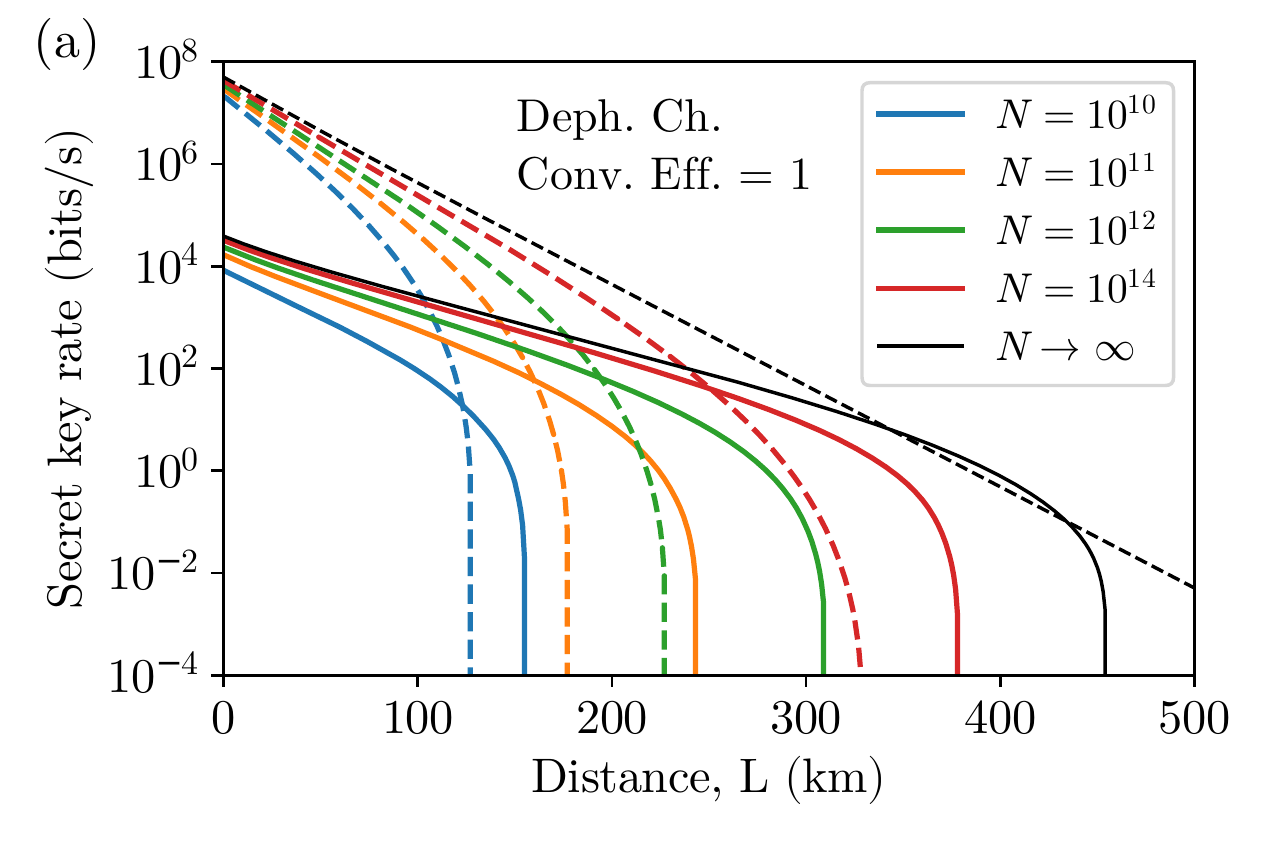}
		\includegraphics[width=0.48\textwidth]{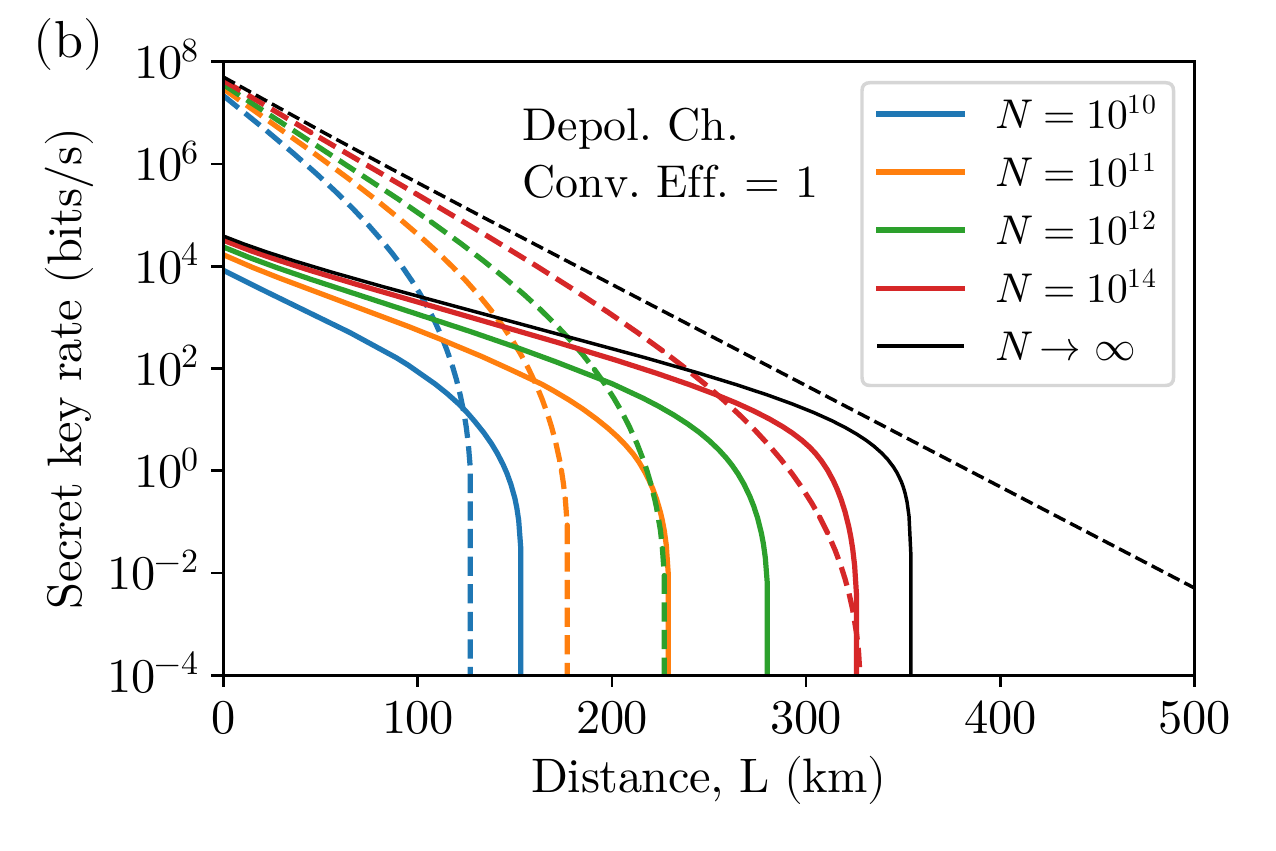}
		\includegraphics[width=0.48\textwidth]{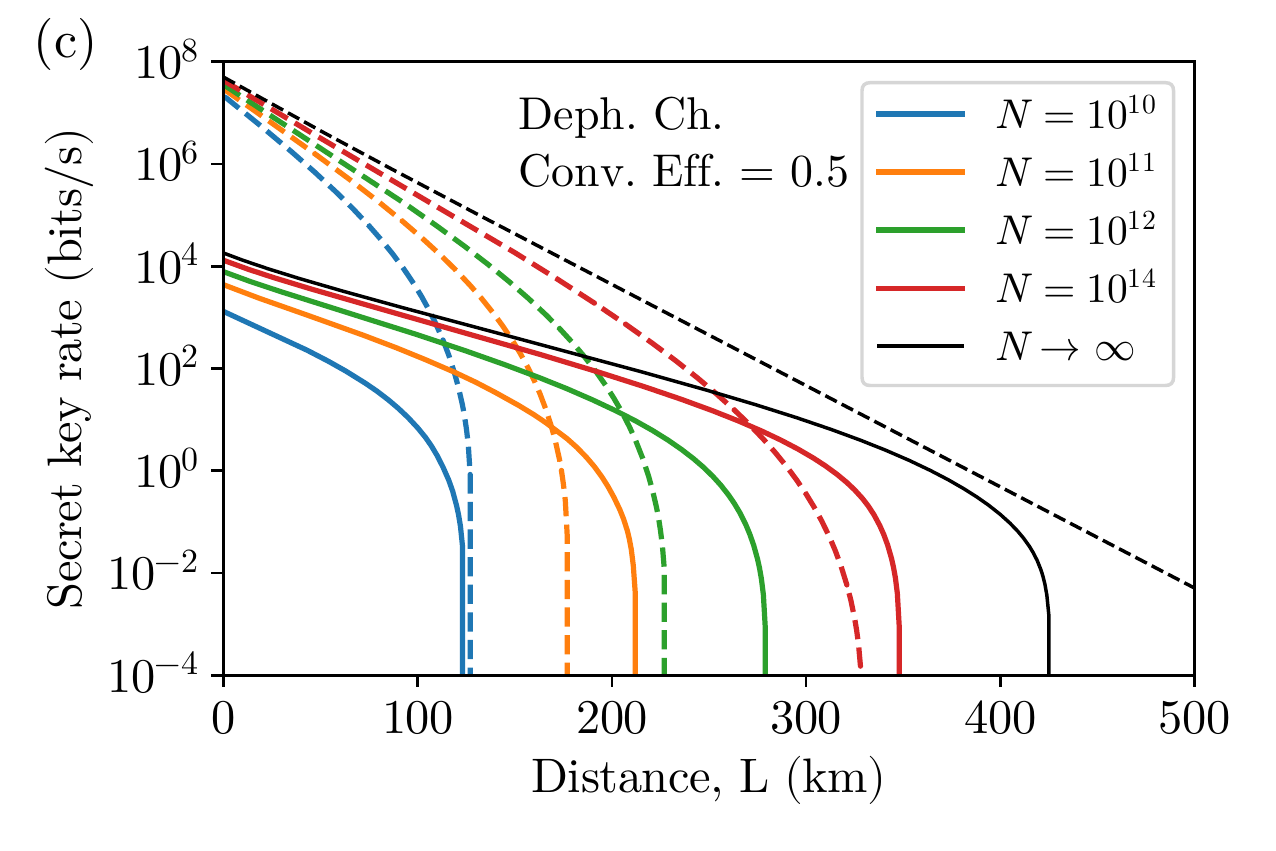}
    		\caption{Secret key generation rate, in b/s, for an MA-QKD setup using warm vapour quantum memories \cite{camacho2009four} (solid lines), in comparison with no-memory MDI-QKD (dashed lines), for different values of the block size $N$. In (a) and (c) a dephasing channel is used to model memory decoherence, whereas, in (b), a depolarisation channel is used. The efficiency of the frequency converter is assumed ideal in (a) and (b), whereas, in (c), it is 50\%. }
		\label{graph:WV_fixblocksize}
	\end{figure}

	In \cref{graph:WV_fixblocksize}, we show the performance of the warm vapour memory in Ref.~\cite{camacho2009four}, for different values of the block size $N$, which represents the total number of signals sent by Alice (or Bob) in that run of the protocol. \red{We can see that, at low distances, the key rate of MA-QKD is lower than that of MDI-QKD. This is partly due to the lower repetition rate for MA-QKD, but also due to the additional loss effects introduced by the QM's less-than-one writing and reading efficiencies. At longer distances, however, the improved key-rate scaling of MA-QKD with channel loss may overcome these effects.
	}
	In \cref{graph:WV_fixblocksize}(a), we can see that in the asymptotic regime (black curves), the MA-QKD protocol can only offer a small advantage over MDI-QKD from around $\SI{340}{\kilo\metre}$ to $\SI{430}{\kilo\metre}$. However, once we use a finite block size $N$ (colour curves),  the crossover distance moves to the left to shorter channel lengths, and even approaches 100~km at $N=10^{10}$. This suggests that in order to see the advantages of MA-QKD over no-QM MDI-QKD we only need to demonstrate such systems over much shorter distances than one may require in the asymptotic regime. With record distances for entanglement distribution between two QMs being around 50~km \cite{EntgDist50km}, one can hope that such a demonstration can take place in near future.
	
	While a slight shift to the left, due to finite-key effects, might be expected in \cref{graph:WV_fixblocksize}, the considerable change in the crossover distance may come as a surprise. A naive thinking may suggest that in order to see the benefits in the finite-key setting, we need to have larger count numbers in MA-QKD, as compared to MDI-QKD, to reduce statistical errors in our parameter estimation. But, so long as, in the asymptotic case, the key rate for MDI-QKD is higher than that of MA-QKD, we may expect that the corresponding counts will also remain larger in the finite-key setting, hence no considerable change may be expected in the crossover distance. This argument, however, fails to give us an accurate picture of what is happening in the MA-QKD case. Below, we explain two key reasons for why the finite-key setting may benefit the MA-QKD setup, hence shifting the crossover distance to much shorter channel lengths.  
	\begin{itemize}
	    \item{\bf Self-purification of multi-photon terms:} The MA-QKD system can by design get rid of some of the erroneous terms that would otherwise be present in the no-QM setup. Let us compare the two setups when Alice selects a non-vacuum intensity $s$, in the $X$ basis, and Bob selects the vacuum intensity $v$. In no-QM MDI-QKD, there is a single BSM module, in which Alice's and Bob's emissions are directly combined. A successful BSM, in polarisation encoding, is declared if two detectors corresponding to different polarisations click. In the event that Bob sends a vacuum state, a successful BSM could happen because of the multi-photon terms in Alice's signal. This increases $M^{sv}$  and $E^{sv}$ counts, which add to the uncertainty in estimating $e_{\rm ph}$. In MA-QKD, such counts are much lower. Charlie will declare that Bob's QM has been loaded when his corresponding side BSM is successful. For a vacuum input, such an event could only happen if one of the detectors clicks because of the dark count, assuming that the EPP source can only cause a click in one of the detectors. For low dark count rates, as we assume here, the measurement counts $M^{sv}$, as well as its corresponding terms in error will be close to zero in MA-QKD. Around the crossover distance, this makes the upper bound on $e_{\rm ph}$ lower for MA-QKD even if its corresponding value in the asymptotic case is higher than that of MDI-QKD. That is, MA-QKD enjoys less noisy statistics that helps us obtain tighter bounds on our parameters of interest.
	
	\item{\bf Efficient use of decoy states:} In both MDI-QKD and MA-QKD, the secret key is extracted from events in which both users select the signal intensity $z$. The rounds in which they both employ the decoy intensities are used for parameter estimation only. The points that one user uses the $Z$ basis and the other uses the $X$ basis, are then somehow "wasted" and will be sifted out. MA-QKD can help with better sifting efficiency. This is partly because of the main advantage of MA-QKD with respect to MDI-QKD in that the key rate scales with the transmissivity of one leg of the channel, rather than the entire channel. To better understand this point, let us consider the effect of employing the vacuum intensity, $v$. Suppose that Alice and Bob are using either an MDI-QKD or an MA-QKD setup with a channel transmittance per leg of $\eta$, and that they use intensity $z$ with probability $p_z \simeq 1$, as they do in the infinite key regime. Charlie will report a successful detection with probability $Q_z$. Now suppose that they use the same scheme as above, except that they now employ a (fictitious) finite-key scheme, in which they employ the vacuum intensity $v$ with probability $p_v = p_z = 1/2$. The effect of this is equivalent to using a channel with transmittance per leg of $\eta/2$, since the effective transmittance of each user's link has been reduced by one half. Since MDI-QKD scales with $\eta^2$, $Q_z$ will be reduced by a factor of 4. However, since MA-QKD scales with ${\eta}$, $Q_z$ will only be reduced by a factor of 2. In reality, Alice and Bob will use additional decoy intensities other than the vacuum intensity. But since the decoy states will typically have larger vacuum components than the signal intensity $z$, they will have a similar effect as adding loss to the system, which MA-QKD tolerates better.
	
	\end{itemize}
	
	Another important factor in our finite-key comparison is the amount of time needed to collect data for a block size $N$. In the case of MDI-QKD, we can typically run the system at a high repetition rate on the order of GHz for very long periods of time. The stability of the memory-based system may, however, require us to stop collecting data after a certain period of time. It would be interesting to see how the two systems compare if, instead of the block size, one fixes the total data collection time $T_{\rm col}$, instead. This corresponds to a block size of $N = R_s T_{\rm col}$, for each system, and gives a considerable advantage to the faster system in collecting more data at an identical time. This would not make much a difference in the case of warm vapours as we can already run the system at sub-GHz rates. But, in the case of cold atomic ensembles or silicon vacancy centres, which represent slower memories, this would be interesting to study. 
	
\begin{figure}[htbp]
		\centering
		\includegraphics[width=0.48\textwidth]{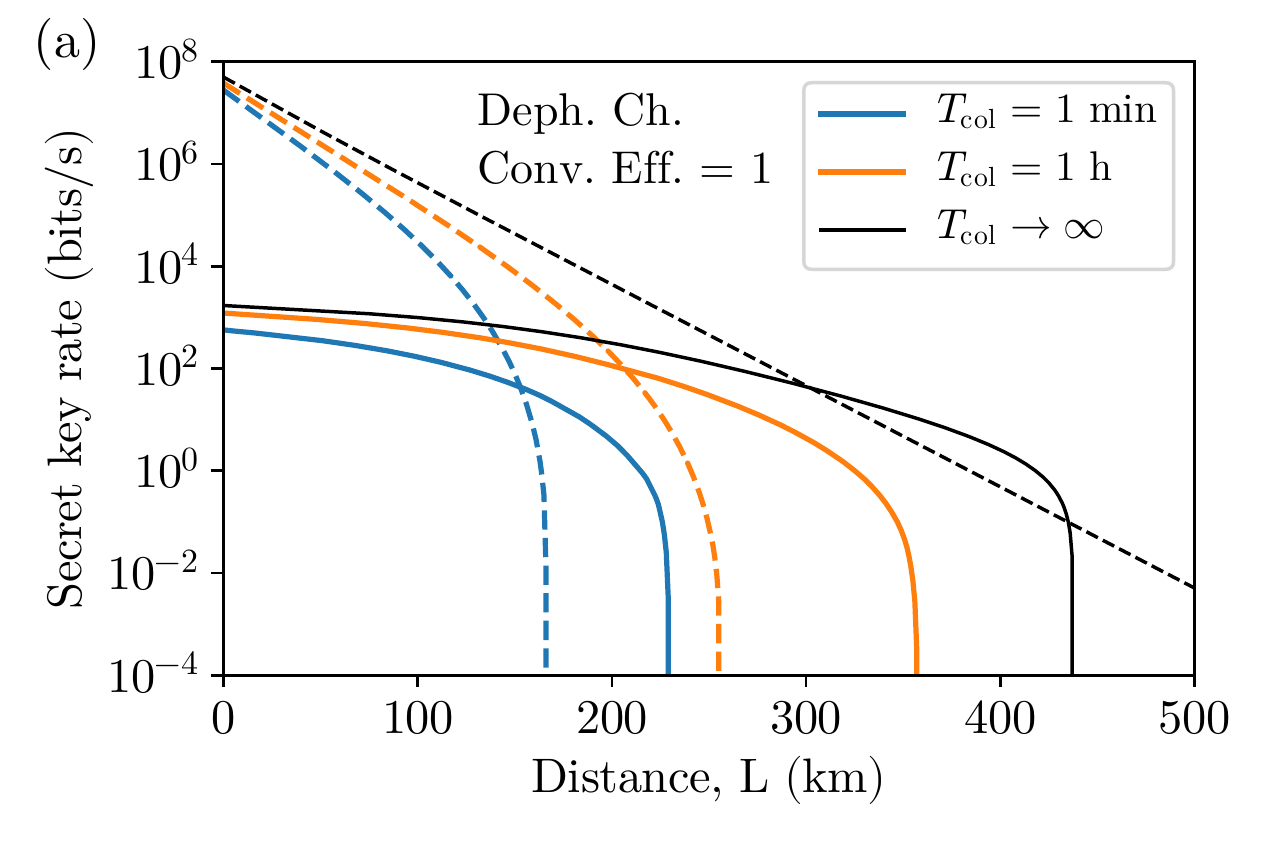}
		\includegraphics[width=0.48\textwidth]{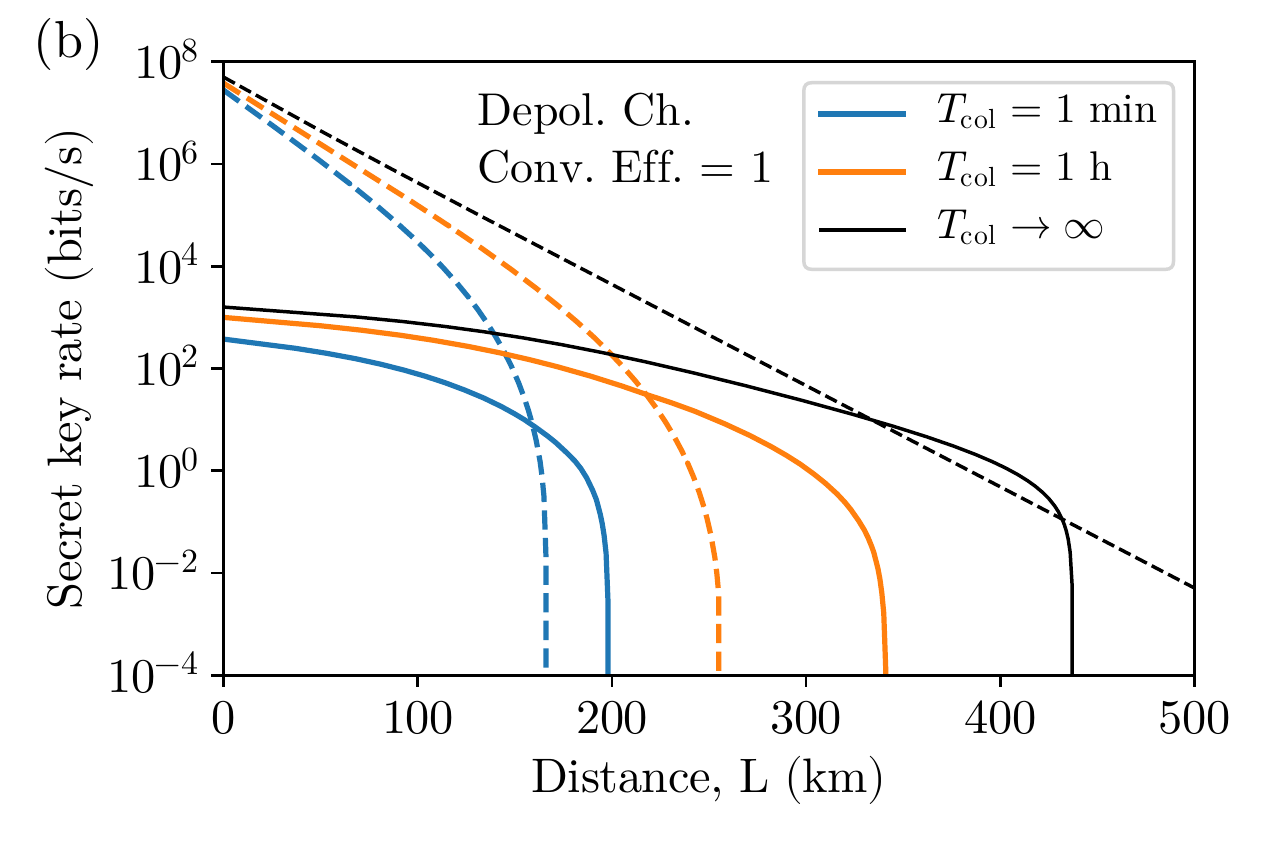}
		\includegraphics[width=0.48\textwidth]{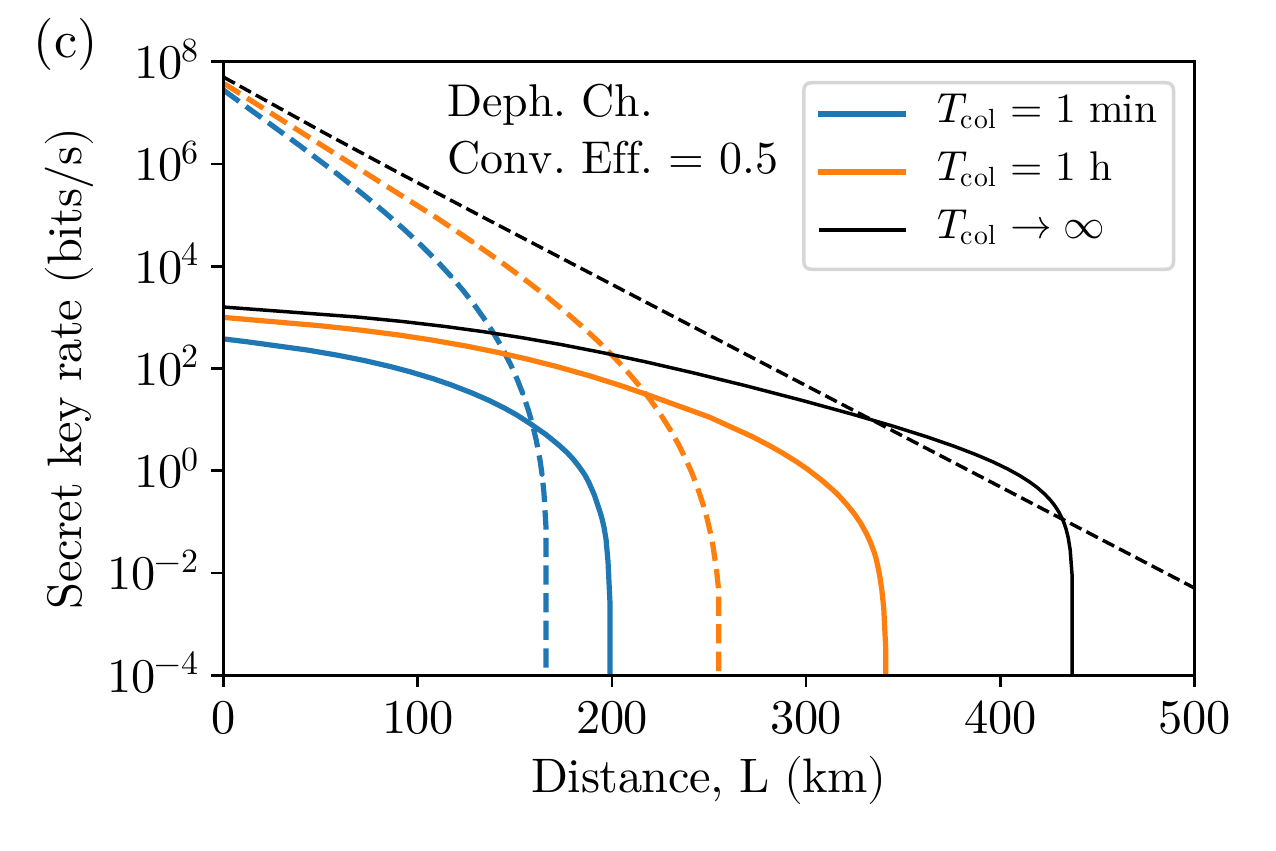}		
    	\includegraphics[width=0.48\textwidth]{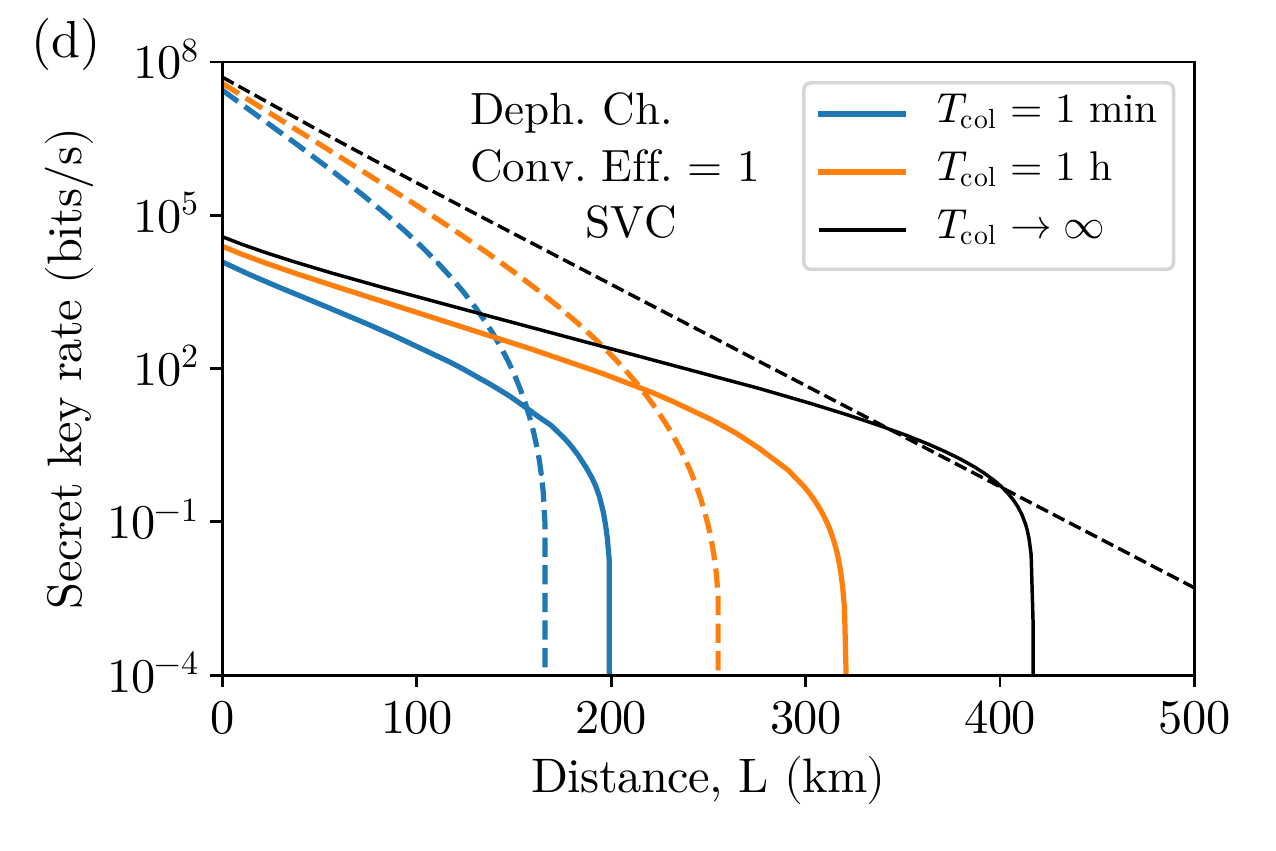}	
		\caption{Secret key generation rate, in b/s, for an MA-QKD setup (solid lines) using (a)--(c) cold atom quantum memories, reported in \cite{yang2016efficient}, and (d) silicon vacancy centres, reported in \cite{bhaskar2020experimental}, in comparison with no-memory MDI-QKD (dashed lines), if we collect data for one minute (blue), one hour (orange), or with no time limit (black). In (a), (c), and (d) a dephasing channel is used to model memory decoherence, whereas, in (b), a depolarisation channel is used. The efficiency of the frequency converter is assumed ideal in (a), (b), and (d), whereas, in (c), it is 50\%.}	\label{graphCA}
	\end{figure}

	Figure \ref{graphCA} \red{(a)-(c)} show the performance of MA-QKD using the cold atom QM reported in Ref.~\cite{yang2016efficient}, with a repetition rate of $\SI{4.2}{\mega\hertz}$, at different collection times. This means that, at an identical collection time, the MDI-QKD system can collect almost 250 times more data than the MA-QKD setup. It is interesting to see that, even under these harsher conditions, the MA-QKD system can offer a similar advantage as we saw in \cref{graph:WV_fixblocksize} over the no-QM MDI-QKD setup. As shown in \cref{graphCA}(a), for a dephasing channel, in the asymptotic regime (black curves), the MA-QKD system can only offer a small advantage in the range from $\SI{300}{\kilo\metre}$ to $\SI{430}{\kilo\metre}$. However, if the experiment is run for an hour (orange curves), MA-QKD can generate more key after $\SI{230}{\kilo\metre}$, and, while MDI-QKD dies off at about $\SI{250}{\kilo\metre}$, MA-QKD can generate a key up to $\SI{350}{\kilo\metre}$. If the experiment is run for just a minute (blue curves), MA-QKD can offer an advantage after a distance of just $\SI{170}{\kilo\metre}$. \red{In \cref{graphCA}(d), we show a similar graph for the silicon vacancy centres used in the recent MA-QKD experiment reported in \cite{bhaskar2020experimental}. This system has a slightly higher repetition rate, but a lower coherence time. The latter is the main reason why the cut-off distance is shorter in \cref{graphCA}(d) compared with  \cref{graphCA}(a).}
		
	\red{Note that it may not be possible to use a memory-based system continuously for a long period of time without applying certain calibrations or cooling techniques. This could reduce the time available for data collection, reducing the effective block size for an MA-QKD system. One key technique that may mitigate this problem in the setup considered in this work is the delayed writing procedure, in which we only attempt to interact with the memory if the corresponding side-BSM is successful. This means that the memory is kept in a ready-to-go initial state until we know a photon has survived the path loss, in which case its state is teleported to the memory. Given that at long distances the chance of the latter event is low, this suggests that the external interaction with the memory is not that frequent, and the time between any two such events can be used to bring the memory back to a solid initial state. In the case of memories reported in \cite{camacho2009four} and \cite{yang2016efficient}, we also have the additional advantage that after reading the memory, it automatically goes back to its initial state. Nevertheless, it is easy in our analysis to consider the effect of possible interruptions in data collection by modifying the block size. For instance, for CA ensembles, we have verified that the advantage shown in \cref{graphCA}(a) will remain even if we can only collect data a quarter of the experiment time.}
	
	Finally, we have looked at how different system parameters can affect the conclusion we draw above. In \cref{graph:WV_fixblocksize}(b) and  \cref{graphCA}(b), we have used a depolarising channel to model the decoherence effect. In comparison to \cref{graph:WV_fixblocksize}(a) and  \cref{graphCA}(a), where a dephasing model is used, we see that the warm vapour system, which has lower $T_2$ values, is more adversely affected than the cold atom system. We observe the same behaviour when we change the frequency converter efficiency from one to to 0.5 as can be seen in \cref{graph:WV_fixblocksize}(c) and  \cref{graphCA}(c). This can simply be a ramification of having noisier data in the case of warm vapours as compared to the cold atom case. This would result in less tight bounds on system parameters at the same block size or collection time, hence sharper drop in key rates. The overall effect would nevertheless suggest that MA-QKD systems can offer competitive performances in the finite-key regime irrespective of the memory or other relevant system parameters. This would be an essential observation in the early demonstrations of memory-based systems and how we benchmark them against their rival counterparts.
	
	


	\section{Conclusions}
	\label{sec-conclusions}
	
	By borrowing ideas from quantum repeaters, MA-QKD can improve the scaling of repeaterless QKD systems. However, the common imperfections in memory-based systems such as their coupling efficiency to photonic systems, or their finite coherence times, may make it difficult for them to offer any practical advantage as compared to their no-memory counterparts. In particular, previous analyses suggest that any advantage in the total key rate would often come only after a crossover distance that is still challenging to implement experimentally. In this work, we showed that once we considered the finite-key effects in the key rate analysis, the crossover distance in such systems was reduced to a point that an experimental implementation could be foreseen in the near future. This effect was attributed to two features of decoy-state MA-QKD systems. First is their ability to purify some of the errors that result from multi-photon terms in weak laser pulses, and the other relates to a more efficient sifting of signal and decoy states. It is essential, however, for MA-QKD systems to keep all sources of noise near the memory units low, as they otherwise would translate into erroneous measurements in the middle site. As such are the multiple excitation terms in the memories, or sources that drive them, or additional background noise that may enter the setup. All these issues are manageable with careful design and they are all precursors to implementing longer quantum communications links relying on quantum memory units. \red{In particular, we believe that the results of this work would be applicable to possible architectures for future quantum networks, in which end users are only equipped with simple equipment, such as decoy-state BB84 encoders, but the core of the network has advanced memory-based repeater chains \cite{piparo2014long}. }
	
	We should note that there are no-memory QKD systems, such as twin-field (TF) QKD \cite{lucamarini2018overcoming}, that offer a similar rate-vs-distance scaling as MA-QKD, and they have already been implemented at record distances \cite{TF-QKD509km}. An MA-QKD system may not be currently able to offer higher key rates or reach longer distances than those achieved by TF-QKD systems. But, it is important to recognise that the expertise and skills in both MA-QKD and TF-QKD would be required to implement scalable quantum repeater systems that go beyond the current rate-versus-distance records. In this respect, this work makes us one step closer to the final goal of implementing long-distance quantum communications systems.
	
	\section*{Acknowledgements}
	
	We thank William Munro and Koji Azuma for valuable discussions. This work was supported by the European Union's Horizon 2020 research and innovation programme under the Marie Sklodowska-Curie grant agreement number 675662 (QCALL). All data generated can be reproduced by the equations and the methodology introduced in this paper.

	\clearpage
	
	\appendix
	
	{\section{Simulation model}
	\label{app-memoryloading}
	
	}

In this appendix, we describe our simulation model, starting with our analysis of the indirect-loading of QMs with attenuated laser sources. {Here, we assume that Charlie is honest, there is no eavesdropper, and we are only interested in finding the relevant parameters in a realistic setting.}

\begin{figure}[h]
	\vspace{10pt}
	\centering
	\includegraphics[width=0.6\textwidth]{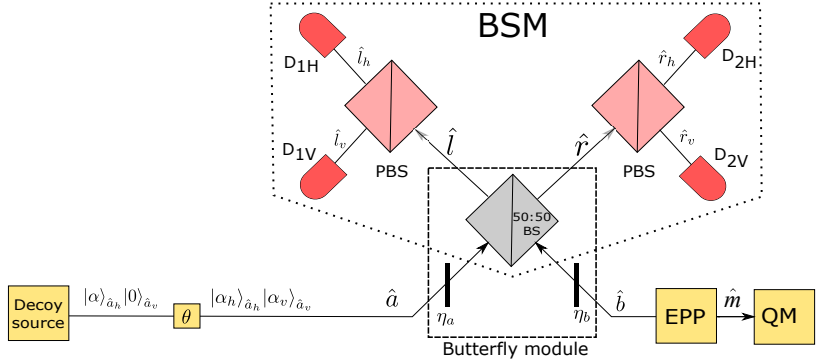}	
	\caption{Loading of a QM with a $Z$-encoded weak coherent pulse, in a round with a misalignment angle of $\theta$. The module in the dotted box represents a partial Bell-state measurement (BSM) on polarisation-encoded photons. We refer to the module in the dashed box as the butterfly module, in which $\eta_a$ models the channel transmissivity and the quantum efficiency of a single-photon detector, whereas $\eta_b$ captures the coupling and frequency conversion efficiencies as well as the quantum efficiency of a single-photon detector. The quantum efficiency of photodetectors in the BSM module is then assumed to be one.}\label{fig:WCPSchematics}
\end{figure}

{Figure \ref{fig:WCPSchematics} shows a schematic view of our memory loading model for a single user, say Alice, in the polarisation encoding case. We model the loss in the channel, the measurement devices, and possible frequency converters as two beam splitters of transmissivity $\eta_a = \eta_{\rm ch} \eta_d$ and $\eta_b = \eta_{c} \eta_d$ located at each input port of the 50:50 beam splitter of the BSM module. Here, $\eta_{\rm ch}$ models the transmissivity of the Alice-Charlie channel, $\eta_{c}$ models the frequency conversion and/or coupling efficiency, and $\eta_d$ represents the efficiency of the single-photon detectors. Note that by assuming the same efficiency $\eta_d$ for all detectors, we are able to analyse its effects at the input ports of the BSM, simplifying our model. We do not need to consider the effect of the QM's writing efficiency, $\eta_w$, at the loading stage. Instead, we modify the reading efficiency $\eta_r$ by an $\eta_w$ factor, allowing us to analyse its effect at the reading stage. In \cref{fig:WCPSchematics}, the EPP source is assumed to generate an ideal entangled state in the form $\tfrac{1}{\sqrt{2}} (\ket{HH}_{\hat{b} \hat{m}} + \ket{VV}_{\hat{b} \hat{m}})$, where $\hat{b}$ and $\hat{m}$, respectively, represent the two output modes of the EPP source heading toward the BSM module and the QM.}

We also consider setup misalignment between the user sources and the central node, which, in polarisation encoding, we model as a random rotation of the horizontal and vertical modes. For simplicity, we assume that the rotation angle $\theta$ is independent and identically distributed between {different} rounds of the protocol, and for the two legs of the system. Also, we assume that polarisation maintenance schemes are in place, so that the reference frames at the user sources and the central node are the same on average. It is reasonable then to assume, as we do in this work, that the probability density function (PDF) $f(\theta)$ is an even function of $\theta$. {One can use a similar formulation when other types of encoding, e.g. time-bin, are used.}

In the following, in Sec.~\ref{Sec:MemLoad}, we first find the post-measurement state of the loaded memory, the loading probability, and the its corresponding error rate under above considerations. The particular issue of misalignment turns out to complicate the analysis when we use weak laser pulses (WCPs) as compared to single-photon sources. Previous analyses of MA-QKD either assume no channel misalignment \cite{abruzzo2014measurement, piparo2017multiple} or model {it} as an error probability $e_{\rm mis}$ \cite{panayi2014memory, piparo2015ensemble}, which is effectively given by $\int_{-\pi}^{\pi}{f(\theta)\sin^2(\theta) {\rm d}\theta}$. In our case, while the analysis is more cumbersome, the end result, in terms of the form of the post-measurement state of the QM, is similar to the single-photon case. This allows us to replicate most of the analysis in \cite{panayi2014memory} in Sec.~\ref{Sec:simulation}, and extend it to the case of depolarisation channels. In the last section of this Appendix, we have summarised the key rate relationships used for the no-QM MDI-QKD as a reference point.

\subsection{Memory loading}
\label{Sec:MemLoad}
Here, we calculate the post-measurement state of the QM, its loading probability and error rate, in the two cases of $Z$ and $X$ bases.

\subsubsection{Analysis for $Z$ basis}
Without loss of generality, let us consider the case that the user generates a horizontally polarised {WCP of intensity $\mu$. Ideally, the state  generated is of the form $\ket{\alpha}_{\hat{a}_h} \ket{0}_{\hat{a}_v}$, where $\alpha = \sqrt{\mu}$} and $\hat{a}_h$ and $\hat{a}_v$ represent, respectively, the horizontal and vertical modes of the transmitted light in \cref{fig:WCPSchematics}. In a particular round with a misalignment angle of $\theta$, the misaligned state, at the input of the butterfly module, is given by
\begin{equation}
\ket{\psi}_{\hat{a}}^{\theta} =  \ket{\alpha_h}_{\hat{a}_h}  \ket{\alpha_v}_{\hat{a}_v}, 
\end{equation}
where $\alpha_h =  \alpha \cos{\theta}$ and $\alpha_v =  \alpha \sin{\theta}$. Meanwhile, the joint state of the two output modes of the EPP source, i.e., $\hat{b}$ and $\hat{m}$, is given by
\begin{equation}
\label{eq:psibmZ}
\ket{\Phi^+}_{\hat{b} \hat{m}} = {\tfrac{1}{\sqrt{2}} (\ket{HH}_{\hat{b} \hat{m}} + \ket{VV}_{\hat{b} \hat{m}})} = \tfrac{1}{\sqrt{2}} (\ket{10H}_{\hat{b}_h \hat{b}_v \hat{m}} + \ket{01V}_{\hat{b}_h \hat{b}_v \hat{m}}),
\end{equation}
where in the last equality, we have divided $\hat{b}$ into, respectively, horizontal and vertical modes $\hat{b}_h$ and $\hat{b}_v$. After reordering modes, and averaging over $\theta$, the joint input state to the butterfly module is given by
\begin{equation}
\label{eq:inpst}
	\hat \rho_{\rm in} = \int_{-\pi}^{\pi} f(\theta) \hat \rho_{\rm in}^\theta {\rm d} \theta ,
\end{equation}
where
\begin{equation}
\label{eq:inpst_theta}
\begin{aligned} 
{\hat \rho_{\rm in}^\theta} = \ket{\psi}_{\hat{a}}^{\theta}\!\Bra{\psi} \otimes \ket{\Phi^+}_{\hat{b} \hat{m}} \! \bra{\Phi^+} &= \frac{1}{2} \Ket{\alpha_h}\!\Bra{\alpha_h}_{\hat{a}_h}  \Ket{1} \! \Bra{1}_{\hat{b}_h}  \Ket{\alpha_v} \! \Bra{\alpha_v}_{\hat{a}_v}  \Ket{0} \! \Bra{0}_{\hat{b}_v}  \Ket{H} \! \Bra{H}_{\hat{m}} \\ 
&+ \frac{1}{2} \Ket{\alpha_h} \! \Bra{\alpha_h}_{\hat{a}_h} \Ket{0} \! \Bra{0}_{\hat{b}_h} \Ket{\alpha_v} \! \Bra{\alpha_v}_{\hat{a}_v} \Ket{1} \! \Bra{1}_{\hat{b}_v} \Ket{V} \! \Bra{V}_{\hat{m}} \\
&+ \frac{1}{2} \Ket{\alpha_h} \! \Bra{\alpha_h}_{\hat{a}_h} \Ket{1} \! \Bra{0}_{\hat{b}_h} \Ket{\alpha_v} \! \Bra{\alpha_v}_{\hat{a}_v} \Ket{0} \! \Bra{1}_{\hat{b}_v} \Ket{H} \! \Bra{V}_{\hat{m}} \\
&+ \frac{1}{2} \Ket{\alpha_h} \! \Bra{\alpha_h}_{\hat{a}_h} \Ket{0} \! \Bra{1}_{\hat{b}_h} \Ket{\alpha_v} \! \Bra{\alpha_v}_{\hat{a}_v} \Ket{1} \! \Bra{0}_{\hat{b}_v} \Ket{V} \! \Bra{H}_{\hat{m}},
\end{aligned}
\end{equation}
and $\ket{\psi}\bra{\psi}_{\hat{a}}$ is our shorthand notation for $\ket{\psi}_{\hat{a}\hat{a}}\!\bra{\psi}$. 

We are interested in the state projected to the QM after a successful loading, i.e., when exactly an H detector and a V detector click in the BSM module. To model this measurement process, we should find the output state of the butterfly module, with an input state as in \cref{eq:inpst}, and then find the post-measurement state for the desired measurement outcome. The key to calculate this is to realise that the horizontal and vertical modes will interact separately at the 50:50 beam splitter of the butterfly module, and will cause clicks in the horizontal and vertically polarised detectors, respectively. Thus, we can split the overall transformation $\hat B$ for the butterfly module in \cref{fig:WCPSchematics}, and the overall POVM operator $\hat M$ in horizontal and vertical operators as follows:
\begin{gather}
\hat B = \hat B_{h} \otimes \hat B_{{v}} \\
\hat M = \hat M_{{h}} \otimes \hat M_{{v}}.
\end{gather}
Here, the {butterfly} operators $\hat B_{{h}}$ and $\hat B_{{v}}$ in \cref{fig:WCPSchematics} only differ in their input and output modes: $\hat B_{{h}}$ will take modes $\hat{a}_h$ and $\hat{b}_h$ to modes $\hat{l}_h$ and $\hat{r}_h$, while $\hat B_{{v}}$ will take modes $\hat{a}_v$ and $\hat{b}_v$ to modes $\hat{l}_v$ and $\hat{r}_v$.
The measurement operators (POVMs) are also identical for both the horizontal and vertical modes, and are given by
\begin{equation}
\begin{aligned}
\hat{M}_{{x}} &= (1-p_{\rm dc}) \left[\left(\hat I_{\hat{l}_x} - (1-p_{\rm dc}) \Ket{0} \! \Bra{0}_{\hat{l}_x}\right) \otimes  \Ket{0} \! \Bra{0}_{\hat{r}_x} \right] \\ 
&+ (1-p_{\rm dc}) \left[\Ket{0} \! \Bra{0}_{\hat{l}_x} \otimes \left(\hat I_{\hat{r}_x} - (1-p_{\rm dc}) \Ket{0} \! \Bra{0}_{\hat{r}_x}\right) \right],
\end{aligned}
\label{eq:measurementop}
\end{equation} 
for $x \in \{h, v\}$, where $\hat I$ is the identity operator for the corresponding mode. $\hat M_{x}$ represents the event of getting a click in the $x$-polarised left detector and no click on the $x$-polarised right detector, or vice-versa.

Using the above notation, the post-measurement state of the QM, after a successful loading, is given by
\begin{equation}
	\label{eq:rhohatmZ}
	\begin{gathered}
	\hat \rho_{\hat{m}} = \frac{\pTr{\hat{l}_h,\hat{l}_v,\hat{r}_h \hat{r}_v}{\hat B^\dagger \hat \rho_{\rm in} \hat B \hat M}}{\Tr \left[\hat B^\dagger \hat \rho_{\rm in}  \hat B \hat M \right]} = \frac{1}{p_{\rm load}^\mu} \int_{-\pi}^{\pi}f(\theta)\pTr{\hat{l}_h,\hat{l}_v,\hat{r}_h \hat{r}_v}{\hat B^\dagger \hat \rho_{\rm in}^\theta \hat B \hat M} {\rm d}\theta
	\end{gathered}
	\end{equation}
	where
		\begin{equation}
		\label{eq:rhomZ1}
	\begin{gathered}
 \pTr{\hat{l}_h,\hat{l}_v,\hat{r}_h \hat{r}_v}{\hat B^\dagger \hat \rho_{\rm in}^\theta \hat B \hat M} =  	c_{HH}(\theta) \ketbra{H} + 	c_{VV}(\theta) \ketbra{V} + c_{HV}(\theta) \ketbra{H}{V} + c_{VH}(\theta) \ketbra{V}{H},
	\end{gathered}
	\end{equation}
	with
	\begin{equation}
	\label{eq:abcdZ}
	\begin{aligned}
	c_{HH}(\theta) &= \frac{1}{2} \Tr \left[\hat B_{h}^\dagger \Ket{\alpha_h} \! \Bra{\alpha_h}_{\hat{a}_h} \Ket{1} \! \Bra{1}_{\hat{b}_h}\hat B_{h} \hat M_{h}\right]
	\Tr \left[ \hat B_{v}^\dagger \Ket{\alpha_v} \! \Bra{\alpha_v}_{\hat{a}_v} \Ket{0} \! \Bra{0}_{\hat{b}_v}\hat B_{v} \hat M_{v}\right] 	 \\
	c_{VV}(\theta) &= \frac{1}{2} \Tr \left[\hat B_{h}^\dagger \Ket{\alpha_h} \! \Bra{\alpha_h}_{\hat{a}_h} \Ket{0} \! \Bra{0}_{\hat{b}_h}\hat B_{h} \hat M_{h}\right] \Tr \left[\hat B_{v}^\dagger \Ket{\alpha_v} \! \Bra{\alpha_v}_{\hat{a}_v} \Ket{1} \! \Bra{1}_{\hat{b}_v}\hat B_{v} \hat M_{v}\right]  \\
	c_{HV}(\theta) &= \frac{1}{2} \Tr \left[\hat B_{h}^\dagger \Ket{\alpha_h} \! \Bra{\alpha_h}_{\hat{a}_h} \Ket{1} \! \Bra{0}_{\hat{b}_h} \hat B_{h}  \hat M_{h}\right] \Tr \left[\hat B_{v}^\dagger \Ket{\alpha_v} \! \Bra{\alpha_v}_{\hat{a}_v} \Ket{0} \! \Bra{1}_{\hat{b}_v} \hat B_{v} \hat M_{v}\right] \\
	c_{VH}(\theta) &= \frac{1}{2} \Tr \left[\hat B_{h}^\dagger \Ket{\alpha_h} \! \Bra{\alpha_h}_{\hat{a}_h} \Ket{0} \! \Bra{1}_{\hat{b}_h} \hat B_{h} \hat M_{h}\right] \Tr \left[\hat B_{v}^\dagger \Ket{\alpha_v} \! \Bra{\alpha_v}_{\hat{a}_v} \Ket{1} \! \Bra{0}_{\hat{b}_v} \hat B_{v} \hat M_{v}\right],
	\end{aligned}
	\end{equation}
and
\begin{equation}
\label{eq:ploadZ}
		p_{\rm load}^\mu = \Tr \left[\hat B^\dagger \hat \rho_{\rm in}  \hat B \hat M \right] = \int_{-\pi}^{\pi} f(\theta) [c_{HH}(\theta)+c_{VV}(\theta)] {\rm d}\theta 
\end{equation} 
is the probability of a successful loading for a WCP with intensity $\mu$. 		%
		
Every individual trace term in \cref{eq:abcdZ} involves either horizontal or vertical modes, and is equivalent to the probability of having exactly one detector click in the corresponding polarisation. Such terms have already been calculated in Table III of \cite{piparo2014long}, which here we reuse, after making necessary adjustments, to obtain	
	\begin{equation}
		\begin{aligned}
		{c_{HH} (\theta)} 
		&=  \left(1-p_{\rm dc} \right) ^{2} \left( 1-{{\rm e}^{-1/2\,\eta_{a}\,
				\left( \sin ^2{\theta} \right){\it \mu}}}
		\left(1-p_{\rm dc} \right)  \right) \times\\
		&  \left(  \left( \eta_{b}\, \left( 
		\cos ^2{\theta} \right){\it \mu}\,\eta_{a}-2\,
		\eta_{b}+4 \right) {{\rm e}^{1/2\,\eta_{a}\, \left( \cos ^2{\theta} \right){\it \mu}}}
		-4\, \left( 1-\eta_{b} \right)
		\left( 1-p_{\rm dc} \right)  \right) {{\rm e}^{-1/2\,\eta_{a}\,{\it \mu}\,
				\left(  \left( \cos^2{\theta} \right)+1
				\right) }}, \\
		c_{VV} (\theta) &=  \left(1-p_{\rm dc} \right) ^{2} \Big[  \left(1-p_{\rm dc} \right) 
 \left( \eta_{b}\,  \cos^2 \theta \mu
\,\eta_{a}-\eta_{b}\,\eta_{a}\,\mu+2\,\eta_{b}-4 \right) {{\rm e}^{-1/
2\,\eta_{a}\,\mu\, \left(  \cos^2 \theta 
+1 \right) }}\\ 
&-4\, \left( 1-\eta_{b} \right)  \left(1-p_{\rm dc}
 \right) {{\rm e}^{1/2\,\eta_{a}\,\mu\, \left(  \cos^2
\theta-2 \right) }}-\left( \eta_{b}\, 
\cos^2 \theta\mu\,\eta_{a}-\eta_{b}\,\eta_{
a}\,\mu+2\,\eta_{b}-4 \right) {{\rm e}^{-1/2\,\eta_{a}\,\mu}}\\
&+4\,{
{\rm e}^{-\eta_{a}\,\mu}} \left( -1+p_{\rm dc} \right) ^{2} \left(1- \eta_{b}
\right)  \Big],
		\end{aligned}
		\end{equation}
and	
\begin{equation}
		c_{HV}(\theta) = c_{VH}(\theta) = \frac{1}{4} \cos \theta \sin \theta (1-p_{\rm dc})^2 (\eta_a \eta_b \mu e^{-\eta_a \mu}).
\end{equation}
It is interesting that, in the above, the diagonal terms $c_{HV}$ and $c_{VH}$ are odd functions of $\theta$. Under our assumption that $f(\theta)$ is an even function, we have that
\begin{equation}
	\int_{-\pi}^{\pi}f(\theta) 	c_{HV}(\theta)  {\rm d}\theta = 	\int_{-\pi}^{\pi}f(\theta) 	c_{VH}(\theta)  {\rm d}\theta = 0,
\end{equation}
implying that these terms vanish when considering the average post-measurement state $\hat \rho_{\hat{m}}$ in \cref{eq:rhohatmZ}. Thus, $\hat \rho_{\hat{m}}$ can be expressed as 
\begin{equation} 
\label{eq:postmeasst}
	\rho_{\hat{m}} = e_{\rm load}^\mu \ketbra{H} + (1-e_{\rm load}^\mu) \ketbra{V},
	\end{equation} 
where
\begin{equation}
\label{eq:eloadZ}
\begin{aligned}
	 e_{\rm load}^\mu &= \frac{1}{p_{\rm load}^\mu} \int_{-\pi}^{\pi} f(\theta) c_{HH} (\theta) {\rm d} \theta 
	 \end{aligned} 
\end{equation}
is the probability of loading the memory with the wrong state. In our case, when we send H-polarised light, a successful BSM in \cref{fig:WCPSchematics} suggests that the $\hat{b}$ mode is V-polarised. The state stored in the memory, for an EPP source with $\ket{\Phi^+}_{\hat{b} \hat{m}}$ as its initial state, is then also expected to be V-polarised. That is why the coefficient for $\ketbra{H}$, in \cref{eq:postmeasst}, represents the loading error probability, in $Z$ basis, for a WCP with intensity $\mu$.

Due to the symmetry of the setup, if the user sends vertically polarised light, the loading probability $p_{\rm load}^\mu$ would be the same, but the post-measurement state is given by $\rho_{\hat{m}} = (1-e_{\rm load}^\mu) \ketbra{H} + e_{\rm load}^\mu \ketbra{V}$.

\subsubsection{Analysis for $X$ basis}

Without loss of generality, let us assume that Alice generates the plus state given by
\begin{equation}
\Ket{\frac{\alpha}{\sqrt{2}}}_{\hat{a}_h} \! \Ket{\frac{\alpha}{\sqrt{2}} }_{\hat{a}_v}.
\end{equation}
In a particular round with a misalignment angle $\theta$, the butterfly module will receive the state
\begin{equation}
\Ket{\psi}_{\hat{a}}^{\theta} =  \Ket{\frac{\alpha}{\sqrt{2}} (\sin{\theta} + \cos{\theta})}_{\hat{a}_h} \! \Ket{\frac{\alpha}{\sqrt{2}} (\sin{\theta} - \cos{\theta})}_{\hat{a}_v},
\end{equation}
while the output state of the EPP source can be written as
\begin{equation}
\Ket{\Phi^+}_{\hat{b} \hat{m}} = \tfrac{1}{\sqrt{2}} (\Ket{DD}_{\hat{b} \hat{m}} + \Ket{AA}_{\hat{b} \hat{m}})
= \tfrac{1}{\sqrt{2}} \left((\Ket{10} + \Ket{01}) \Ket{D} + (\Ket{10} - \Ket{01}) \Ket{A})\right)_{\hat{b}_h \hat{b}_v \hat{m}},
\end{equation}
where $\ket{D} = (\ket{H} + \ket{V})/\sqrt{2}$ and $\ket{A} = (\ket{H} - \ket{V})/\sqrt{2}$.

The analysis is similar to the one for the $Z$ basis. After going through similar steps, we find that the probability to successfully load the memory is given by
{\begin{equation}
	\label{eq:ploadX}
	\begin{aligned}
	{p_{\rm load}^\mu} &= \int_{-\pi}^{\pi} f(\theta) \frac{1}{2}\, \left(1-p_{\rm dc} \right) ^{2} \Bigl(   \left(1-p_{\rm dc} \right)  \left( \cos \theta\sin
	\left( \theta \right) \mu\eta_{a}\,\eta_{b}-1/2\,\eta_{b}\,\mu\eta_{a}+6\,\eta_{b}-8 \right) {{\rm e}^{-1/2\,\eta_{a}\,\mu\left( \cos \theta\sin \left( \theta
			\right) +3/2 \right) }} \\
	&- \left(1-p_{\rm dc} \right) 
	\left( \cos \theta\sin \left( \theta \right) \mu\eta_{a}\,\eta_{b}+1/2\,\eta_{b}\,\mu\eta_{a}-6\,\eta_{b}
	+8 \right) {{\rm e}^{1/4\,\eta_{a}\,\mu \left( 2\,\cos
			\left( \theta \right) \sin \left( \theta \right) -3 \right) }} \\
	&+ \left( \eta_{b}\,\mu\eta_{a}-4\,
	\eta_{b}+8 \right) {{\rm e}^{-1/2\,\eta_{a}\,\mu}}+8\,{
		{\rm e}^{-\eta_{a}\,\mu}} \left(1-p_{\rm dc} \right) ^{2}
	\left(1-\eta_{b} \right)  \Bigr) {\rm d} \theta,
	\end{aligned}
	\end{equation}}
and, under our assumption that $f(\theta)$ is even, the post-measurement state of the memory can be written as
\begin{equation} \label{eq:postmeasstX}
\rho_{\hat{m}} = e_{\rm load}^\mu \ketbra{D} + (1-e_{\rm load}^\mu) \ketbra{A},
\end{equation} 
where
{\begin{equation}
	\label{eq:eloadX}
	\begin{aligned}
	e_{\rm load}^\mu &= \frac{1}{p_{\rm load}^\mu} \int_{-\pi}^{\pi} f(\theta) \frac{1}{4}\, \left( -1+p_{\rm dc} \right) ^{2} \Bigl(    \left(1-p_{\rm dc} \right)  \left( \cos \left( {
		\it \theta} \right) \sin \left( {\it \theta} \right) \mu\eta_{a
	}\,\eta_{b}-1/2\,\eta_{b}\,\mu\eta_{a}+6\,\eta_{b}-8 \right) 
	{{\rm e}^{-1/2\,\eta_{a}\,\mu \left( \cos \left( {\it \theta}
			\right) \sin \left( {\it \theta} \right) +3/2 \right) }}\\
	&- \left(1-p_{\rm dc} \right) 
	\left( \cos \left( {\it \theta} \right) \sin \left( {\it \theta}
	\right) \mu\eta_{a}\,\eta_{b}+1/2\,\eta_{b}\,\mu
	\eta_{a}-6\,\eta_{b}+8 \right) {{\rm e}^{1/4\,\eta_{a}\,\mu
			\left( 2\,\cos \left( {\it \theta} \right) \sin \left( {\it \theta}
			\right) -3 \right) }} \\
	&+ \left(2\,\eta_{b}\,\mu\eta_{a} -2\,
	\left( \cos^2{\theta} \right)\mu \eta_
	{a}\,\eta_{b}-4\,\eta_{b}+8 \right) 
	{{\rm e}^{-1/2\,\eta_{a}\,\mu}}+8\,{{\rm e}^{-\eta_{a}\,\mu}} \left(1-p_{\rm dc} \right) ^{2} \left(1-\eta_{b} 
	\right) \Bigr) {\rm d}\theta.
	\end{aligned}
	\end{equation}}

Finally, note that we calculate the integrals in \cref{eq:ploadZ,eq:eloadZ,eq:ploadX,eq:eloadX} numerically as a closed form expression for them could not be found. In our simulations, to compute $p_{\rm load}^\mu$ and $e_{\rm load}^\mu$, we assume that $f(\theta)$ follows a uniform distribution over $[-\Theta, \Theta]$. To have a fair comparison with no-memory MDI-QKD, we choose $\Theta = \sqrt{3 e_{\rm mis}}$, where $e_{\rm mis}$ is the misalignment error probability in one leg of a symmetric MDI-QKD setup. This is motivated by the fact that
	\begin{equation}
	\label{eq:misalign-equiv}
		\frac{1}{2\sqrt{3 e_{\rm mis}}} \int_{-\sqrt{3 e_{\rm mis}}}^{\sqrt{3 e_{\rm mis}}} \sin^2 \theta {\rm d}\theta \approx \frac{1}{2\sqrt{3 e_{\rm mis}}}	\int_{-\sqrt{3 e_{\rm mis}}}^{\sqrt{3 e_{\rm mis}}} \theta^2 {\rm d}\theta = e_{\rm mis},
	\end{equation}
	which implies that the chosen $f(\theta)$ would cause a misalignment error of approximately $e_{\rm mis}$ in the MDI-QKD setup.%
	
	\subsection{Key rate simulation}
	\label{Sec:simulation}
	In Sec.~\ref{Sec:MemLoad}, we showed that the post-measurement QM state after a successful loading is a mixture of the desired and undesired states for the QM; see \cref{eq:postmeasst} and \cref{eq:postmeasstX}. In effect, it is as if the state of QM has flipped with a probability $e_{\rm load}^\mu$. This is similar to how misalignment acts on a single photon state, because of which  we can think of the whole loading process as a channel with an effective misalignment of $e_{\rm load}^\mu$. This would also make it possible to use the methodology in Ref.~\cite{panayi2014memory} to calculate the required parameters of the key rate formula. In particular, the photonic states retrieved from the two QMs turn out to also have a similar form to a misaligned photon, although at a higher error rate to account for the dephasing/depolarisation process.
	
	
	In the following, we explain how to simulate all terms in the key-rate formula, in both the asymptotic and finite-key regimes. Given that in MA-QKD, one of the memories will be read immediately after loading, only one of the QMs would undergo the decay process. That implies that the middle BSM in \cref{fig:schematics} can be thought as an asymmetric MDI-QKD setup, with possibly different transmissivities $\eta_l$ and $\eta_r$ for, respectively, its left and right legs \cite{panayi2014memory}. We can then use the yield and error rate formulas, summarised below, of asymmetric single-photon MDI-QKD for our rate calculation:
	\begin{gather}
	\label{eq:Y11_MDI}
		 Y_{11}^{\rm MDI} (\eta_l, \eta_r) = (1-p_d)^2 \left[\frac{\eta_l \eta_r}{2}+(2 \eta_l + 2 \eta_r - 3 \eta_l \eta_r) p_d + 4(1-\eta_l) (1-\eta_r) p_d^2\right], \\ 
		 e_{11;X}^{\rm MDI} (\eta_l,\eta_r,e_d)  Y_{11}^{\rm MDI} (\eta_l, \eta_r) = e_0   Y_{11}^{\rm MDI} (\eta_l, \eta_r)-(e_0-e_d) (1-p_d)^2 \eta_l \eta_r /2, \\
		 e_{11;Z}^{\rm MDI} (\eta_l,\eta_r,e_d)  Y_{11}^{\rm MDI} (\eta_l, \eta_r) = e_0  Y_{11}^{\rm MDI} (\eta_l, \eta_r)-(e_0-e_d) (1-p_d)^2 (1-2 p_d) \eta_l \eta_r /2,
	\end{gather} 
	where $e_0 = 1/2$ and $e_d$ is the total misalignment probability in the asymmetric MDI-QKD setup, i.e., the probability that exactly one of the photons is misaligned. 
	
	\subsubsection{Asymptotic regime}
	\label{sec:AppAsymp}
	In this case, the key-rate formula is given by \cref{eq:keyrateasymp}. In this regime, we assume that the signal intensity $z$, encoded in the $Z$-basis, is chosen with probability approaching one, and the parameter estimation provides perfect estimates of the single-photon terms $Q_{11}^{Z}$ and $e_{\rm ph}$. We only then need to simulate the values of $Q_Z$, $e_Z$, $Q_{11}^{Z}$ and $e_{\rm ph}$ under nominal mode of operation. The procedure we use to calculate these terms is very similar to that of \cite{panayi2014memory}. The main differences are our new model for the memory-loading with WCPs, developed earlier in this Appendix, and the inclusion of the depolarising channel for memory decoherence.
	
	To compute $Q_Z$, we divide it into two parts: (1) the probability of having the two memories loaded and available to read in a given round, denoted by $P_{\rm side}$, and (2) the probability that the middle BSM is successful, given that the QMs are ready, denoted by $P_{\rm mid}$. Then,
	\begin{equation}
	\label{eq:QZ}
	Q_Z = P_{\rm side} P_{\rm mid}. 
	\end{equation}
	To find $P_{\rm side}$, we first estimate the probability to load the QM with a Z-encoded WCP, given by $p_{\rm load}^z$ in \cref{eq:ploadZ}. Then, we compute the average number of rounds $N_L$ that it takes to load both memories, substituting $\eta_A$ and $\eta_B$ by $p_{\rm load}^z$ in Eq.~(C.3) of \cite{panayi2014memory}, to obtain
	\begin{equation}
	N_L = \frac{3 - 2 p_{\rm load}^z}{p_{\rm load}^z  (2 - p_{\rm load}^z)}.
	\end{equation}
	Then, we have that
	\begin{equation}
	P_{\rm side} = \frac{1}{N_L + N_r}.
	\end{equation}
	where $N_r$ is the number of rounds it takes to read the memory, which we assume to be one.
	
	The second term is given by
	\begin{equation}
	P_{\rm mid} = Y_{11}^{\rm MDI} (\eta_m,\eta_{m'}),
	\end{equation}
	where $\eta_m = \eta_w \eta_{r0} \eta_d$ is the effective reading efficiency of the QM loaded later, and $\eta_{m'}$ is the average effective reading efficiency of the QM loaded earlier, given by \cite{panayi2014memory}
	\begin{equation}
	\eta_{m'} = \frac{(1+e^{T/T_1}-p_{\rm load}^z)p_{\rm load}^z }{(2-p_{\rm load}^z)(e^{T/T_1}+p_{\rm load}^z-1)} \eta_m,
	\end{equation}
	where $T_1$ is the time constant for the decay process of the QM.
		
	The single-photon component $Q_{11}^Z$ is given by
	\begin{equation}
	Q_{11}^Z = Q_Z \frac{(p_{\rm load}^{\rm SP})^2}{(p_{\rm load}^z)^2} z^2 e^{-2 z},
	\end{equation}
	where $p_{\rm load}^{\rm SP}$ is the probability to load the QM when a single photon is sent, given by \cite{panayi2014memory}
	\begin{equation}
	\label{eq:ploadSP}
	p_{\rm load}^{\rm SP} =  Y_{11}^{\rm MDI} (\eta_{ch} \eta_d,\eta_c \eta_d).
	\end{equation}
	
	To find $e_{\rm ph}$, we first calculate the misalignment-error probability for loading the QM with an $X$-basis single photon, which is given by \cite{panayi2014memory}
	\begin{equation}
	e_{\rm load}^{X,\rm SP} =  e_{11;X}^{\rm MDI} (\eta_{ch} \eta_d,\eta_c \eta_d, e_{\rm mis}).
	\end{equation}
	Then, we obtain
	\begin{equation}
		e_{\rm ph} =  e_{11;X}^{\rm MDI} (\eta_m,\eta_m',\avg{e_{\rm QM}^{\rm SP}}),
	\end{equation}
	where $\avg{e_{\rm QM}^{\rm SP}}$ is the total misalignment probability, given by
	\begin{equation}
		\label{eq:eQmSPdeph}
\avg{e_{\rm QM}^{\rm SP}} = 	2 e_{\rm load}^{X,\rm SP} + 2 \beta  \avg{e_{\rm deph}} -2 	e_{\rm load}^{X,\rm SP} e_{\rm load}^{X,\rm SP} -4 \beta  \avg{e_{\rm deph}} e_{\rm load}^{X,\rm SP},
	\end{equation}
	with
	\begin{equation}
	\label{eq:edephSPS}
 \avg{e_{\rm deph}} = 1 - \frac{p_{\rm load}^z}{1-(1-p_{\rm load}^z)^2} - \frac{(p_{\rm load}^z)^2 (1-p_{\rm load}^z e^{-T/T_2})}{[1-(1-p_{\rm load}^z) e^{-T/T_2}][1-(1-p_{\rm load}^z)^2]},
	\end{equation}
	in the case of dephasing memories, and by
	\begin{equation}
	\avg{e_{\rm QM}^{\rm SP}} = 	2 e_{\rm load}^{X,\rm SP} + 2 \beta \avg{e_{\rm depol}} -2 	e_{\rm load}^{X,\rm SP} e_{\rm load}^{X,\rm SP} -4 \beta \avg{e_{\rm depol}} e_{\rm load}^{X,\rm SP},
\end{equation}
with
	\begin{equation}
	\label{eq:edepolSPS}
\avg{e_{\rm depol}} = \frac{2}{3} \avg{e_{\rm deph}},
\end{equation}
in the case of depolarising memories. 

	To calculate $e_Z$, we use
	\begin{equation}
		e_Z =  e_{11;Z}^{\rm MDI} (\eta_m,\eta_m',\textrm{E} \left\{e_{\rm QM}\right\}),
	\end{equation}
	where $\textrm{E}\left\{e_{\rm QM}\right\}$ is the average total misalignment-error probability between the two QMs, which depends on the specific model used for decoherence. In the dephasing model, the $Z$-basis QM states will not be affected by the decoherence, therefore, the probability that exactly one state is misaligned is as follows
	\begin{equation}
		\textrm{E}\left\{e_{\rm QM}\right\}= e_{\rm QM} = 2 e_{\rm load}^z (1-e_{\rm load}^z),
	\end{equation}
	where $e_{\rm load}^z$ is given by \cref{eq:eloadZ}. For the depolarisation model, we have
	\begin{equation}
	\label{eq:eQM}
\avg{e_{\rm QM}} =  2 e_{\rm load}^z  +2 \beta \avg{e_{\rm depol}} -2 e_{\rm load}^z e_{\rm load}^z  - 4 \beta \avg{e_{\rm depol}} e_{\rm load}^z,
\end{equation}
where $\beta = 1-2 e_{\rm load}^z$.

To derive \cref{eq:eQM} and Eqs.~(\ref{eq:eQmSPdeph})~to~(\ref{eq:edepolSPS}), we have used a similar analysis as in Appendix D of Ref.~\cite{panayi2014memory}.

	\subsubsection{Finite-key regime}
	\label{Sec:AppFinite}
	
	In this case, we need to calculate the sets $\{M^{ab}\}$ and $\{E^{ab}\}$, where $M^{ab}$ is the total number of measurement counts when Alice (Bob) has used intensity $a$ ($b$), while $E^{ab}$ is the number of such events that also result in an error. Note that intensity $z$ is encoded in the $Z$ basis and intensities $\{w_1,w_2,v\}$ are encoded in the $X$ basis; we are only interested in estimating $\{M^{ab}\}$ and $\{E^{ab}\}$ when $a,b$ are encoded in the same basis.
	
	For our numerical simulations, we still need to make some assumptions on the obtained measurement results in a nominal experiment. For this purpose, we use the expected values for relevant parameters using the corresponding probability in the asymptotic regime. That is, we assume
	\begin{equation}
	\label{eq:eab}
	M^{ab} = N  Q^{ab} \mbox{\ \ and \ \ }
	E^{ab} = e_{ab} M^{ab},
	\end{equation}
	where $N$ is the total number of rounds, i.e., the number of transmitted pulses by Alice/Bob, in the protocol, $Q^{ab}$ is the probability of having a successful measurement originating from intensities $a$, for Alice, and $b$, for Bob, and $e_{ab}$ is the probability that this measurement results in an error.
	
	To calculate $Q_{ab}$, we first compute the total gain $Q_{\rm tot}$, using the same procedure as for $Q_Z$ in the asymptotic case, with the difference that $Q_{\rm tot}$ is now a function of the average memory-loading probability given by
	\begin{equation}
	\bar{p}_{\rm load} =  \sum_{a} p_a p_{\rm load}^a,
	\end{equation}
	where $p_{a}$ is the probability of selecting intensity $a \in \{z, w_1, w_2, v\}$; and $p_{\rm load}^a$ is the probability of a successful loading when the user selects intensity $a$, given by either \cref{eq:eloadZ} or \cref{eq:eloadX}, depending on whether intensity $a$ is encoded in the $Z$ or $X$ basis.
	Then, we have that
	\begin{gather}
	N_L = \frac{3 - 2 	\bar{p}_{\rm load}}{	\bar{p}_{\rm load}  (2 - \bar{p}_{\rm load})}, \\
	\eta_{m'} = \frac{(1+e^{T/T_1}-\bar{p}_{\rm load})\bar{p}_{\rm load} }{(2-\bar{p}_{\rm load})(e^{T/T_1}+\bar{p}_{\rm load}-1)} \eta_m, \\
	P_{\rm side} = \frac{1}{N_L + N_r}\\
	P_{\rm mid} = Y_{11}^{\rm MDI} (\eta_m,\eta_{m'}), \\
	Q_{\rm tot} = P_{\rm side} P_{\rm mid},
	\end{gather}
	where $N_r = 1$ and  $\eta_m = \eta_w \eta_{r0} \eta_d$. Now, $Q^{ab}$ is the fraction of $Q_{\rm tot}$ that originated from intensities $a,b$. Note that after a successful loading, the state projected to the QM is always a misaligned qubit. The probability that the middle BSM is successful only depends on the loss coefficients $\eta_{m}$ and $\eta_{m'}$, and it is independent of the intensities $a,b$ that caused the loading. Thus, $Q^{ab}$ only depends on how likely intensities $a,b$ are to cause a successful loading, that is,
	\begin{equation}
		Q^{ab} = Q_{\rm tot} p_a p_b \frac{p_{\rm load}^a p_{\rm load}^b}{\bar{p}_{\rm load}^2}.
	\end{equation}

	For $e_{ab}$, we have that
	\begin{gather}
	e_{zz} =  e_{11;Z}^{\rm MDI} (\eta_m,\eta_m',\avg{e_{zz}^{\rm QM}}), \\
	e_{ab} =  e_{11;X}^{\rm MDI} \left(\eta_m,\eta_m',\avg{e_{ab}^{\rm QM}}\right), \quad a,b \in \{w_1, w_2, v\}
	\end{gather}
	where $\avg{e_{ab}^{\rm QM}}$ is the total average misalignment error probability between the two QMs, and depends on whether one considers a dephasing or depolarisation model. The former has no effect on $Z$-basis states, and therefore
	\begin{equation}
		\avg{e_{zz}^{\rm QM}} = e_{zz}^{\rm QM} = 2 e_{\rm load}^z  (1-e_{\rm load}^z). 
	\end{equation}
	For the $X$-basis intensities, we have that
		\begin{equation}
	\avg{e_{ab}^{\rm QM}} = e_{\rm load}^a + e_{\rm load}^b + \beta_a \avg{e_{\rm deph}} + \beta_b \avg{e_{\rm deph}} -2 e_{\rm load}^a e_{\rm load}^b -2 \beta_a \avg{e_{\rm deph}} e_{\rm load}^b - 2 \beta_b \avg{e_{\rm deph}} e_{\rm load}^a,
	\end{equation}
	where $\beta_k = 1-2 e_{\rm load}^k$, and
	\begin{equation}
		\avg{e_{\rm deph}} = 1 - \frac{\bar{p}_{\rm load}}{1-(1-\bar{p}_{\rm load})^2} - \frac{\bar{p}_{\rm load}^2 (1-\bar{p}_{\rm load} e^{-T/T_2})}{[1-(1-\bar{p}_{\rm load}) e^{-T/T_2}][1-(1-\bar{p}_{\rm load})^2]},
	\end{equation}
	using a similar analysis to the one that results in Eq.~(D.8) of \cite{panayi2014memory}.
	
	For a depolarisation channel, we have that, for all intensities
		\begin{equation}
	\avg{e_{ab}^{\rm QM}} = e_{\rm load}^a + e_{\rm load}^b + \beta_a \avg{e_{\rm depol}} + \beta_b \avg{e_{\rm depol}} -2 e_{\rm load}^a e_{\rm load}^b -2 \beta_a \avg{e_{\rm depol}} e_{\rm load}^b - 2 \beta_b \avg{e_{\rm depol}} e_{\rm load}^a,
	\end{equation}
	where
		\begin{equation}
	\avg{e_{\rm depol}} = \frac{2}{3} \avg{e_{\rm deph}}.
	\end{equation}	
	
	\subsection{MDI-QKD without QMs}
	\label{Sec:AppMDI}
	Here, we give the formulas that we have used to simulate the no-memory MDI-QKD with WCP sources.
	
   	In general, if Alice and Bob encode in the $Z$ basis and choose intensities $a$ and $b$, respectively, the gain and error-rate formulas are given by \cite{ma2012statistical}
\begin{gather} \label{eq:Qab_MDIZ}
Q^{ab} = Q_c + Q_e, \\ \label{eq:eab_MDIZ}
e_{ab} = e_d Q_c + (1-e_d) Q_e,
\end{gather}
where $e_d$ represents the total misalignment error probability given by $e_d = 2 e_{\rm mis} {(1-e_{\rm mis}})$, and
\begin{equation}
\begin{gathered}
Q_c = 2(1-p_d)^2 e^{-\zeta/2} (1-(1-p_d) e^{-\eta a/2}) (1-(1-p_d) e^{-\eta b/2}) \\
Q_e = 2 p_d (1-p_d)^2 e^{-\zeta/2}[I_0(2x)-(1-p_d) e^{-\zeta/2} ] \\
x = \eta \sqrt{ab}/2\\
\zeta = \eta (a+b),
\end{gathered}
\end{equation}
where $I_0$ is the modified Bessel function of the first kind and {$\eta = \eta_{\rm ch} \eta_d$} is the total attenuation between each user and the middle node. If they encode in the $X$ basis, they are given by \cite{ma2012statistical}
\begin{gather} \label{eq:Qab_MDIX}
Q^{ab} = 2 y^2 [1+2y^2-4y I_0(x)+I_0(2x)], \\ \label{eq:eab_MDIX}
e_{ab} = \frac{Q^{ab}}{2}-(1-2 e_d) y^2 [I_0(2x)-1],
\end{gather}
where
\begin{equation}
y = (1-p_d)e^{-\zeta/4}.
\end{equation}

\subsubsection{Asymptotic regime}
In the asymptotic regime, the key rate formula is given by
\begin{equation}
R \leq R_s \left[Q_{11}^{Z} \left(1-h(e_{\rm ph})\right) - fQ_{Z} h(e_{Z})  \right].
\end{equation}
$Q_Z$ and $e_Z$ are given by \cref{eq:Qab_MDIZ,eq:eab_MDIZ}, respectively, by substituting $a=b=z$. In the asymptotic regime, we assume that the users are able to obtain perfect estimates of $Q_{11}^{Z}$ and $e_{\rm ph}$, which are given by
\begin{gather} 
Q_{11}^{Z} = z^2 e^{-2z} Y_{11}, \\
e_{\rm ph} =  e_{11;X}^{\rm MDI} (\eta,\eta,e_d) =  \frac{1}{2} - \frac{1}{Y_{11}} (1/2 - e_d) (1-p_d)^2 (1-2 p_d) \frac{\eta^2}{2},
\end{gather}
where
\begin{equation}
Y_{11} =  Y_{11}^{\rm MDI} (\eta, \eta) =  (1-p_d)^2 \left[\frac{\eta^2}{2}+(4 \eta-3\eta^2) p_d+4(1-\eta)^2 p_d^2 \right].
\end{equation}

\subsubsection{Finite-key regime}

We need to simulate the sets $\{M^{ab}\}$ and $\{E^{ab}\}$. In our simulations, we assume that all measurement counts equal their expected values, that is,
\begin{equation}
M^{ab} = N p_{ab} Q^{ab} \mbox{\ \ and \ \ }
E^{ab} = e_{ab} M^{ab},
\end{equation}
where $Q_{ab}$ and $e_{ab}$ are given by \cref{eq:Qab_MDIZ} and \cref{eq:eab_MDIZ} for $Z$-encoded intensities, and by \cref{eq:Qab_MDIX} and \cref{eq:eab_MDIX} for $X$-encoded intensities, and $p_{ab}$ is the probability that Alice and Bob choose intensities $a$ and $b$, respectively.
	
	\section{Finite-key analysis}
	\label{app-finitekey}
	
	In this Appendix, we explain the detailed procedure for finding a lower bound on $M_{11}^Z$ and an upper bound on $e_{\rm ph}$ in \cref{eq:finitekeyrate}. For our finite-key analysis of MDI-QKD and MA-QKD, we use the analytical estimation procedure introduced in \cite{curty2014finite}, together with the tighter multiplicative Chernoff bounds introduced in \cite{zhang2017improved}. Also, as in \cite{zhou2016making}, we estimate the total single photon measurement counts $M_{11}$ in both bases using data in the $X$ basis only. We then link it with $M^{zz}_{11}$ via random sampling analysis. This allows us to encode decoy intensities in the $X$ basis only, thus wasting fewer rounds for statistical estimation.
		
	\subsection{Background}
	\label{sub:finite-intro}
	In the protocol, Alice and Bob emit phase-randomised coherent states of a random intensity $a \in \{z, w_1, w_2, v\}$, where the $z$ intensity is encoded in the $Z$ basis and the rest of the intensities are encoded in the $X$ basis. Without knowing the basis information, the output state corresponding to intensity $a$ can be written as
	\begin{equation}
	\rho_{a} = \sum_{n=0}^{\infty} p_{n|a} \ketbra{n},
	\end{equation}
	where $p_{n|a}$ is the probability that a pulse of intensity $a$ contains $n$ photons, and $\ket{n}$ is the $n$-photon Fock state. For weak laser pulses, we can typically assume a Poisson distribution for the photon number, in which case, $p_{n|a} = a^n e^{-a}/n!$. While most of our analysis does not depend on the choice of the probability distribution, we also use the Poisson assumption for our numerical results. Based on the above diagonal form, for a pulse encoded in a given basis, the only information available to Eve is its photon number $n$. This implies that, instead of the actual protocol, Alice and Bob could have run the equivalent {\em virtual} scenario in which
		\begin{itemize}
		\item Alice (Bob) sends a $Z$-encoded $n$-photon Fock state with probability $p_{n,Z} = p_{z} p_{n|z}$.
		\item Alice (Bob) sends an $X$-encoded $n$-photon Fock state with probability $p_{n,X} = \sum_{a \in \{w_1, w_2, v\}} p_{a} p_{n|a}$.
	\end{itemize}
	
	In this virtual scenario, Alice and Bob can wait until after Eve's attack to assign each emission of an $X$-encoded $n$-photon Fock state to intensity $a \in \{w_1, w_2, v\}$ with probability
	\begin{equation}
		p_{a|n,X} = \frac{p_a p_{n|a}}{p_{n,X}},
	\end{equation}
	and then "reveal" their intensity choices in the appropriate step of the protocol, so that Eve cannot tell which scenario (actual or virtual) is being performed.
	
	Note that Fock states encoded in different bases are in general partially distinguishable to Eve, so Alice and Bob must decide their encoding basis before their emission, even in the virtual scenario. There is one important exception, however: single-photon signals encoded in either the $X$ or $Z$ bases are indistinguishable once averaged by their selection probabilities, since
	\begin{equation}
	\rho_1 = \frac{1}{2}\ketbra{H}{H} + \frac{1}{2}\ketbra{V}{V} = \frac{1}{2}\ketbra{D}{D} + \frac{1}{2}\ketbra{A}{A}.
	\end{equation}
	This implies that the users could have replaced their single-photon emissions by the following purification of $\rho_1$
	\begin{equation}
	\label{eq:delayedbasis}
	\ket{\psi_{1}} = \frac{1}{\sqrt{2}} \left(\ket{0}\ket{H} + \ket{1}\ket{V}\right) = \frac{1}{\sqrt{2}} \left(\ket{+}\ket{D} + \ket{-}\ket{A}\right),
	\end{equation}
	where the first qubit, in $\ket{0}$-$\ket{1}$ basis, is held by the users and $\ket{\pm} = \frac{1}{\sqrt{2}} (\ket{0} \pm \ket{1})$. This allows us to alter our virtual scenario in the following way: when Alice and Bob both decide to send a single-photon state, they replace their respective emissions by the generation of $\ket{\psi_{1}}$, and then wait until after Eve's attack to decide in which basis to measure their ancilla. This delayed basis choice will allow us to estimate the statistics of $Z$-encoded single-photon emissions using $X$-basis data.
	
	\subsection{Estimation of $M_{11}^{Z}$}
	\label{subsec:M11Z}
	
	The estimation is divided in two steps:
	\begin{enumerate}
		\item Estimation of $M_{11}$, the total single-photon measurement counts in both basis, using the decoy state analysis.
		\item Estimation of $M_{11}^Z$ from $M_{11}$, via a random sampling analysis.
	\end{enumerate}
	
	\subsubsection{Estimation of $M_{11}$}
	
	In our virtual scenario, the users have replaced their decoy-state emissions by Fock states, which are only assigned to a particular intensity after Eve's attack. Let $\mathcal{M}_{nm}^X$, with $(n,m) \neq (1,1)$, be the set of rounds in which Alice (Bob) chooses the $X$ basis, sends $n$ ($m$) photons, and Charlie reports a successful detection. Also, let $M_{nm}^X = \abs{\mathcal{M}_{nm}^X}$. After her reports, Alice and Bob will assign each event in $\mathcal{M}_{nm}^X$ to intensities $a,b \in \{w_1, w_2, v\}$ with probability
	
	\begin{equation}
	\label{eq:p_ab_nmX}
	p_{ab|nm,X} = p_{a|n,X} p_{b|m,X} = \frac{p_{a} p_{n|a}}{p_{n,X}} \frac{p_{b} p_{m|b}}{p_{m,X}},
	\end{equation}
	where $p_{n,X} = \sum_{a \in \{w_1, w_2, v\}} p_{a} p_{n|a}$ by the law of total probability. As explained above, Alice and Bob have also delayed their choice of basis on those rounds in which both sent a single photon. Let $\mathcal{M}_{11}$ be the set of rounds in which Alice and Bob sends a single photon and Charlie reports a successful detection, and let $M_{11} = \abs{\mathcal{M}_{11}}$. The probability that they assign each event in $\mathcal{M}_{11}$ to intensities $a,b \in \{z, w_1, w_2, v\}$ is
	\begin{equation}
	\label{eq:p_ab_nm}
	p_{ab|11} = p_{a|1} p_{b|1} = \frac{p_{a} p_{1|a}}{p_1} \frac{p_{b} p_{1|b}}{p_1} 
	\end{equation}
	where $p_{1} = \sum_{a\in \{z,w_1, w_2, v\}} p_{a} p_{n|a}$ by the law of total probability. Let $M^{ab}$ denote the number of rounds assigned to intensities $a,b \in \{w_1, w_2, v\}$. Its expected value is
	\begin{equation}
	\label{eq:Mab2}
	{\rm E}[M^{ab}] = p_{ab|00,X} M_{00}^X + p_{ab|01,X} M_{01}^X + p_{ab|11} M_{11} + \sum_{(m,n)\in S}{p_{ab|mn,X}M_{mn}^X},
	\end{equation}
	where $S = \{(m,n)|m,n \in \mathbb{Z}, m,n \geq 0\} - \{(0,0),(0,1),(1,1)\}$. Each of these intensity assignments is a Bernoulli random variable, and therefore ${\rm E}[M^{ab}]$ is the average value of the sum of some Bernoulli random variables. The values of $M^{ab}$ measured by Alice and Bob correspond to an instance of this sum of Bernoulli random variables. 
	
	Let $\chi = \sum_{i=1}^{n} \chi_i$ be the outcome of the sum of $n$ independent Bernoulli random variables $\chi_i \in \{0,1\}$. Given the observation of the outcome $\chi$, its expectation value $ E[\chi]$ can be bounded by \cite{zhang2017improved}
	\begin{equation}
	\label{eq:chernoff-begin}
	\begin{gathered}
	{\rm E}^{\rm L}[\chi] = \frac{\chi}{1+\delta^{\rm L}}, \\
	{\rm E}^{\rm U}[\chi] = \frac{\chi}{1-\delta^{\rm U}}, 
	\end{gathered}
	\end{equation}
	except with probability $\epsilon$, where $\delta^{\rm L}$ and $\delta^{\rm U}$ are the solutions of the equations
	\begin{equation}
	\label{eq:chernoff-eqs}
	\begin{gathered}
	\left[\frac{e^{\delta^{\rm L}}}{(1+\delta^{\rm L})^{1+\delta^{\rm L}}}\right]^{\chi/(1+\delta^{\rm L})} = \frac{1}{2} \varepsilon \\
	\left[\frac{e^{-\delta^{\rm U}}}{(1-\delta^{\rm U})^{1-\delta^{\rm U}}}\right]^{\chi/(1-\delta^{\rm U})} = \frac{1}{2} \varepsilon.
	\end{gathered}
	\end{equation}
	These solutions can be expressed in terms of the Lambert W function, the inverse of $f(z) = z e^z$, as follows
	\begin{equation}
	\label{eq:lambert}
	\begin{gathered}
	\delta^{\rm L} =  W_0(-e^{\ln(\varepsilon/2-\chi)}/\chi) \\
	\delta^{\rm U} = W_{-1}(-e^{\ln(\varepsilon/2-\chi)}/\chi),
	\end{gathered}
	\end{equation}
	which is useful for their quick numerical computation.
	
	We use \cref{eq:chernoff-begin} to find bounds on $E[M^{ab}]$, which by \cref{eq:Mab2} will set constraints on the values of $M_{nm}^X$ and $M_{11}$. Since we are interested in $M_{11}^{\rm L}$, our analysis can be reformulated as the optimization problem: Find $\min M_{11}$ such that
	
	\begin{equation}
	\label{eq:M11_linearproblem}
	{\rm E}^{\rm L}[M^{ab}] \leq p_{ab|00,X} M_{00}^X + p_{ab|01,X} M_{01}^X + p_{ab|11} M_{11} + \sum_{(m,n)\in S}{p_{ab|mn,X}M_{mn}^X} \leq {\rm E}^{\rm U}[M^{ab}]
	\end{equation}
	$\forall a,b \in \{w_1,w_2,v\}$. 
	This problem can be solved using linear optimisation techniques \cite{curty2014finite}. In this work, however, we use the computationally faster analytical estimation method laid out in the Supplementary Note 1 of \cite{curty2014finite}, for Poisson distributed input signals. {Note that to use this analytical method, one needs to define the term $\hat{M}_{11}^{X}$ such that
	\begin{equation}
	\label{eq:hatM11}
	    p_{ab|11} M_{11} =  p_{ab|11,X} \hat{M}_{11}^{X},
	\end{equation}
	where $p_{ab|11,X}$ is given by \cref{eq:p_ab_nmX}, and substitute $p_{ab|11} M_{11}$ by $p_{ab|11,X} \hat{M}_{11}^{X}$ in \cref{eq:M11_linearproblem}. Then, one can use the results of \cite{curty2014finite} to find a lower bound on $\hat{M}_{11}^{X}$, and reuse \cref{eq:hatM11} to turn it into a lower bound $M_{11}^{\rm L}$ on $M_{11}$.}

%
%
%
%
%
	
	\subsubsection{Estimation of $M_{11}^Z$ from $M_{11}$}
	\label{subsubsec:M11Z}
	
	Let $\mathcal{M}_{11}^Z$ be the subset of $\mathcal{M}_{11}$ in which both users employ the $Z$ basis, and let $M_{11}^Z = \abs{\mathcal{M}_{11}^Z}$. By the delayed basis argument, Alice and Bob could decide which events in $\mathcal{M}_{11}$ belong to $\mathcal{M}_{11}^Z$ after Eve's attack. They assign each event in $\mathcal{M}_{11}$ to $\mathcal{M}_{11}^Z$ with probability
	\begin{equation}
	\label{eq:p_zz_11}
	p_{zz|11} = \left(\frac{p_z p_{1|z}}{p_{1}}\right)^2.
	\end{equation}
	Let $\chi = \sum_{i=1}^{n} \chi_i$ be the outcome of the sum of $n$ independent Bernoulli random variables $\chi_i \in \{0,1\}$. Given the expectation value $E[\chi]$, the outcome $\chi$ can be lower-bounded by \cite{zhang2017improved}
	\begin{equation}
	\label{eq:symmetrical_chernoff}
	\begin{gathered}
	\chi \geq \chi^{\rm L} = (1-\delta) \bar{\chi} \\
	\delta = \frac{-\ln(\varepsilon) + \sqrt{[\ln(\varepsilon)]^2} - 8 \ln(\varepsilon) \bar{\chi}}{2 \bar{\chi}},
	\end{gathered}
	\end{equation}
	except with probability $\varepsilon$. 
	
	The lower bound on $M_{11}^Z$ is then given by $(M_{11}^Z)^L = (1-\delta) \bar \chi $, where $\bar \chi = p_{zz|11} M_{11}^{\rm L}$ and $\delta$ is given by \cref{eq:symmetrical_chernoff}.
		
	\subsection{Estimation of $e_{\rm ph}$}
	
	The upper bound on $e_{\rm ph}$ is given by
	
	\begin{equation}
	e_{\rm ph}^{\rm U} = \frac{(E_{11}^{Z})^{\rm U}}{(M_{11}^Z)^{\rm L}},
	\end{equation}
	where $E_{11}^{Z}$ is the number of phase errors in $\mathcal{M}_{11}^Z$, that is, the number of bit errors that Alice and Bob would have obtained if they had encoded their $Z$ basis single-photon emissions in the $X$ basis. The estimation of this quantity is divided in two steps:
	\begin{enumerate}
		\item Estimation of $E_{11}$, the total amount of phase-flip errors in all single-photon emissions.
		\item Estimation of $E_{11}^{Z}$ from $E_{11}$, via a random sampling analysis.
	\end{enumerate}
		
	\subsubsection{Estimation of $E_{11}$}
	
	Let us imagine that, in the virtual scenario, Alice and Bob measure all their pairs of ancillas in $\mathcal{M}_{11}$ in the X basis, even those that they have assigned to $\mathcal{M}_{11}^Z$. Let $\mathcal{E}_{11}$ be the subset of $\mathcal{M}_{11}$ in which they find a phase-flip error, and let ${E}_{11} = \abs{\mathcal{E}_{11}}$. Each event in $\mathcal{E}_{11}$ is assigned to intensity $a,b \in \{z, w_1, w_2, v\}$ with probability $p_{ab|11}$ defined in \cref{eq:p_ab_nm}.
	
	Also, let $\mathcal{E}_{nm}^X$, with $(n,m) \neq (1,1)$, be the subset of $\mathcal{M}_{nm}^X$ in which Alice and Bob obtain a phase-flip error. Each event in $\mathcal{E}_{nm}^X$ is assigned to intensity $a,b \in \{w_1, w_2, v\}$ with probability $p_{ab|nm,X}$ defined in \cref{eq:p_ab_nmX}. For $a,b \in \{w_1, w_2, v\}$, the expected value of $E_{ab}$ with respect to these assignments is
	\begin{equation}
	\label{eq:EMab2}
	{\rm E}[E^{ab}] = p_{ab|00,X} E_{00}^X + p_{ab|01,X} E_{01}^X + p_{ab|11} E_{11} + \sum_{(m,n)\in S}{p_{ab|mn,X}E_{mn}^X}.
	\end{equation}
	%
%
%
%
	%
	From Eqs. (\ref{eq:chernoff-begin})--(\ref{eq:lambert}), we obtain bounds ${\rm E}^{\rm L}[E^{ab}],{\rm E}^{\rm U}[E^{ab}]$, and redefine our analysis as the optimization problem: Find max $E_{11}$ such that

	\begin{equation}
	\label{eq:E11_linearproblem}
	{\rm E}^{\rm L}[E^{ab}] \leq p_{ab|00,X} E_{00}^X + p_{ab|01,X} E_{01}^X + p_{ab|11} E_{11} + \sum_{(m,n)\in S}{p_{ab|mn,X}E_{mn}^X} \leq {\rm E}^{\rm U}[E^{ab}],
	\end{equation}
	$\forall a,b \in \{w_1,w_2,v\}$. Again, this problem can be solved using linear programming techniques, but we use the analytical estimation method in the Supplementary Note 1 of \cite{curty2014finite}. {Note that to use this analytical method, one needs to define a term $\hat{E}_{11}^{X}$ such that
	\begin{equation}
	\label{eq:hatE11}
	    p_{ab|11} E_{11} =  p_{ab|11,X} \hat{E}_{11}^{X},
	\end{equation}
	where $p_{ab|11,X}$ is given by \cref{eq:p_ab_nmX}, and substitute $p_{ab|11} E_{11}$ by $p_{ab|11,X} \hat{E}_{11}^{X}$ in \cref{eq:E11_linearproblem}. Then, one can use the results of \cite{curty2014finite} to find an upper bound on $\hat{E}_{11}^{X}$, and reuse \cref{eq:hatE11} to turn it into an upper bound $E_{11}^{\rm U}$ on $E_{11}$.}
	
	%
	%
	%
	%
	%
	%
	%
	%
	%
	%
	%
	%
	
	\subsubsection{Estimation of $E_{11}^{Z}$ from $E_{11}$}
	
	By the delayed basis argument, each event in $E_{11}$ will be assigned to $E_{11}^{Z}$ with probability $p_{zz|11}$, defined in \cref{eq:p_zz_11}. 
	
	Let $\chi = \sum_{i=1}^{n} \chi_i$ be the outcome of the sum of $n$ independent Bernoulli random variables $\chi_i \in \{0,1\}$. Given the expectation value $E[\chi]$, the outcome $\chi$ can be upper-bounded by \cite{zhang2017improved}
	\begin{equation}
	\label{eq:symmetrical_chernoff2}
	\begin{gathered}
	\chi \leq \chi^{\rm U} = (1+\delta) \bar{\chi} \\
	\delta = \frac{-\ln(\varepsilon) + \sqrt{[\ln(\varepsilon)]^2} - 8 \ln(\varepsilon) \bar{\chi}}{2 \bar{\chi}},
	\end{gathered}
	\end{equation}
	except with probability $\varepsilon$.
	
	Finally, an upper bound on $E_{11}^{Z}$ is given by $(E_{11}^{Z})^{\rm U} = (1+\delta) \bar \chi $, where $\bar \chi = p_{zz|11} E_{11}^{\rm U}$ and $\delta$ is given by \cref{eq:symmetrical_chernoff2}.

	\bibliography{references}
	
\end{document}